\documentclass[reprint,aps,prx,showpacs,showkeys,groupedaddress,longbibliography]{revtex4-1}

\usepackage{graphicx}        
\usepackage{grffile}         
\usepackage{graphics}
\usepackage{epsfig}
\usepackage{verbatim}
\usepackage{dcolumn}        
\usepackage{bm}            
\usepackage{amsmath}
\usepackage{mathtools}
\usepackage{physics}
\usepackage{hyperref}        
\hypersetup{colorlinks,breaklinks,
	linkcolor=red,urlcolor=red,
	anchorcolor=green,citecolor=blue}
\usepackage{color}
\usepackage{float}
\usepackage{mathrsfs}
\usepackage{braket}
\usepackage{amssymb}
\usepackage{txfonts}
\usepackage{float}
\usepackage{enumitem}
\usepackage{multirow}
\usepackage{orcidlink}

\begin{document}
\title{Stronger Entanglement Dies Faster:  Quantum Mpemba Effect in Dissipative Qubits}
\author{Zhilong Liu~\orcidlink{0009-0000-0353-5113}}
\affiliation{Department of Physics, Key Laboratory of Low Dimensional Quantum Structures and Quantum Control of Ministry of Education, and Synergetic Innovation Center for Quantum Effects and Applications, Hunan Normal University, Changsha, Hunan 410081, P. R. China}
\author{Zehua Tian}
\email{tzh@hznu.edu.cn (Corresponding author)} 
\affiliation{School of Physics, Hangzhou Normal University, Hangzhou, Zhejiang 311121, China}
\author{Jieci Wang~\orcidlink{0000-0001-5072-3096}}
\email{jcwang@hunnu.edu.cn (Corresponding author)}
\affiliation{Department of Physics, Key Laboratory of Low Dimensional Quantum Structures and Quantum Control of Ministry of Education, and Synergetic Innovation Center for Quantum Effects and Applications, Hunan Normal University, Changsha, Hunan 410081, P. R. China}
\begin{abstract}
	 In classical thermodynamics, the Mpemba effect refers to the counterintuitive observation that hot water can freeze faster than cold water, manifesting as an anomalous crossing of dynamical trajectories. While analogues of this phenomenon have been explored in open quantum systems and spin-chain entanglement asymmetry, its connection to the finite-time loss of entanglement remains elusive. In this work, we report a distinct Mpemba effect for quantum entanglement  in a dissipative quantum system associated with entanglement sudden death (ESD). By analyzing two qubits interacting with local amplitude damping reservoirs, we demonstrate that a more strongly entangled initial state can experience a faster collapse into a separable state than a more weakly entangled state. This anomalous decay stems from the competition between initial coherence and excited-state population, where the latter acts as a catalyst for ESD. We provide exact analytical derivations for the trajectory crossover and ESD time, and map the phase diagram to precisely identify the parameter regime where the effect occurs. Our results offer a new strategy for controlling the lifetime of quantum resources in dissipative environments.
\end{abstract}

\maketitle

\emph{Introduction-} 
	The Mpemba effect, a counterintuitive relaxation phenomenon where a system initially further from equilibrium (``hotter") reaches the target state faster than one closer to it (``cooler"), has reemerged as a focal point in non-equilibrium thermodynamics. While its origins trace back to Aristotle and its scientific revitalization began with Mpemba’s 1960s experiments on freezing mixtures~\cite{EBMpemba:1969}, it lacked a rigorous theoretical foundation until recently. In a landmark analysis of classical systems, the Mpemba effect was linked to the nonmonotonic dependence of the initial slow-mode amplitude on the initial temperature. The same framework also predicted the inverse Mpemba effect~\cite{doi:10.1073/pnas.1701264114}. Following this paradigm, investigations have rapidly expanded to anomalous relaxation phenomena in open quantum systems~\cite{Beato:2025dbm,Bao:2025ocj,Nava:2025fdd,Carollo:2021hew,Longhi:2024omn,VanVu:2024zps,Bao:2022ffs,Wang:2024jlr,Strachan:2024mva,Moroder:2024jjh,Chatterjee:2023lgx,Chattopadhyay:2026qgl}, gaining validation across diverse experimental platforms~\cite{PhysRevE.99.060901,Kumar2020ExponentiallyFC,AharonyShapira:2024nrt,Zhang:2024juc}(for a comprehensive review, see Ref.~\cite{Ares:2025onj,Teza:2025azr,bechhoefer2021fresh,Yu:2025vth}). Furthermore, Chang  {\it et al.} extended this anomalous relaxation framework to closed quantum systems utilizing the imaginary-time evolution approach~\cite{Chang:2024ysp,Yu:2025bgy}.
	\begin{figure}[b]
		\centering{\includegraphics[width=0.89\linewidth]{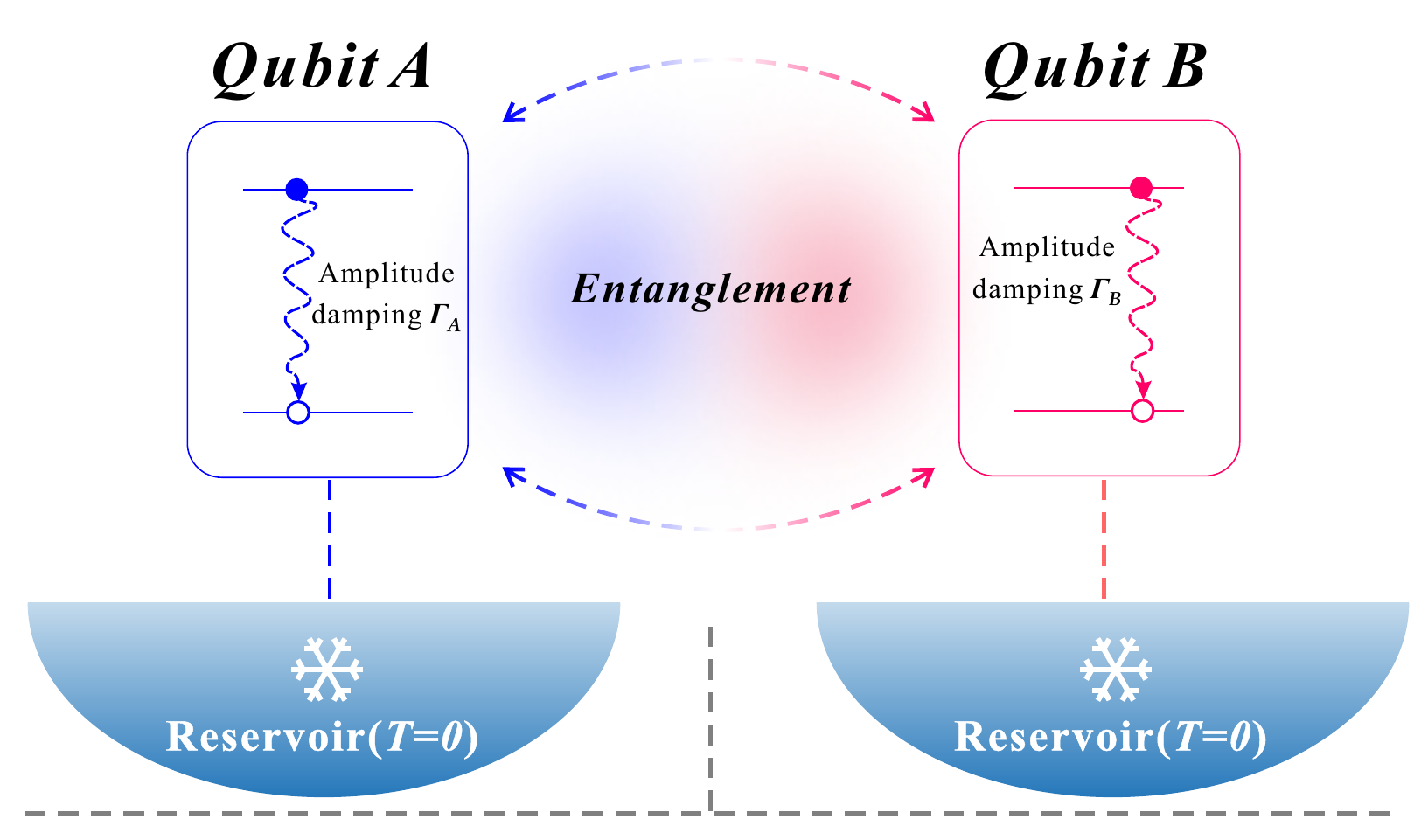}
			\caption{Schematic representation of the dissipative model. A bipartite system, consisting of two-level subsystems $A$ and $B$, is initially prepared in an entangled state and subsequently undergoes local dissipation through independent zero-temperature reservoirs. Each system undergoes an amplitude damping process with a decay rate of $\Gamma_A$ and $\Gamma_B$, respectively. Both systems are coupled to their own independent reservoirs maintained at zero temperature ($T=0$).}\label{fig_SCH}
		}
	\end{figure}
	
	The extension of this effect to the quantum regime has recently revealed novel manifestations that transcend classical interpretations. An entanglement asymmetry Mpemba effect~\cite{Ares:2022koq}, for example, has been observed in closed quantum systems, specifically tilted ferromagnetic states of the XX chain, which subsequently stimulated extensive theoretical~\cite{Yu:2025lku,Klobas:2024png,Summer:2025wsa,Liu:2024uqf,Foligno:2024jpq,Bertini:2022srv,Yamashika:2024hpr,Murciano:2023qrv,Yu:2025omg,Rylands:2023yzx} and experimental research~\cite{Xu:2025wml,Rath:2022qif,Joshi:2024sup,Turkeshi:2024juo,Liu:2024kzv}. Unlike classical thermal relaxation, this purely quantum phenomenon arises from the restoration of broken symmetries. Further developments include a spin wave analysis~\cite{Yamashika:2025vpd} and the Ergotropic Mpemba Effect in Gaussian quantum batteries~\cite{Medina:2024yep}, the latter of which has been extended to other platforms~\cite{Sapui:2026nle,5xrr-x2rm}. These advances highlight the growing importance of understanding how quantum resources exhibit anomalous decay.

	While the Mpemba effect has attracted considerable attention in quantum research, studies on quantum correlations, which play a key role in quantum information and experimental tasks, remain scarce. Whether the dissipative dynamics of quantum correlations can exhibit the Mpemba effect is not only a novel question; the sensitivity of this effect to initial conditions may also provide guidance for the initialization of experiments in practical quantum tasks.
	In practice,  interaction with the environment, whether via thermal reservoirs, electromagnetic noise, or even relativistic effects in spacetime,  inevitably leads to decoherence and entanglement degradation~\cite{Yu:2006iux,Yu:2004cga,Almeida:2007jib,Li:2025bzd,Liu:2025hcx,Gallock-Yoshimura:2021xsy}. This decay limits the operational lifetime of quantum devices and poses a significant hurdle for long-distance quantum tasks. Crucially, the dynamics of this degradation are sensitive not only to the environment but also to the geometric and algebraic properties of the initial state.
	
	In this work, we report an anomalous decay of quantum resources in a dissipative quantum system. We demonstrate a  Mpemba effect for quantum entanglement: an initially more entangled (``hot") state decays faster and reaches ESD (the ``frozen" state), sooner than an initially less entangled (``cold") state. By providing an exact solution to the crossover dynamics, we construct a comprehensive phase diagram that delineates the boundaries of this effect. Our results reveal that the Mpemba effect in quantum systems is not merely a thermal analogy but a fundamental feature of dissipative entanglement evolution, offering potential strategies for optimizing the lifetime of quantum resources.
	
\emph{Models-}
We consider a composite system consisting of two independent two-level subsystems (e.g., atoms or superconducting qubits), labeled $A$ and $B$. As illustrated in Fig.~\ref{fig_SCH}, each subsystem is weakly coupled to its own local bosonic reservoir. We assume the subsystems are spatially separated such that they do not interact directly, and no particle exchange occurs between them. Under the Born-Markov and rotating-wave approximations (RWA), the joint dynamics are described by a quantum master equation. Given that the reservoirs are maintained at zero temperature ($T=0$), they act as purely dissipative sinks that preclude thermal excitations. Consequently, each subsystem undergoes a pure amplitude damping process, physically corresponding to spontaneous emission.

To focus on the non-unitary dissipative dynamics, we adopt the interaction picture with respect to the free system Hamiltonian
\begin{equation}
H_S = \frac{\omega_A}{2}\sigma_z^A + \frac{\omega_B}{2}\sigma_z^B,
\end{equation}
where $\omega_{A(B)}$ is the transition frequency of subsystem $A (B)$. In this frame, the free evolution is eliminated, and the density matrix $\rho(t)$ of the composite system evolves according to the Lindblad master equation (setting $\hbar = 1$):
	\begin{equation}\label{eq_lind}
		\frac{d\rho(t)}{dt}=\mathcal{L}\rho=\sum_{i=A,B}\Gamma_i \left(\sigma_i^-\rho \sigma_i^+ -\frac{1}{2}\left\{\sigma_i^+ \sigma_i^- ,\rho\right\}\right).	
	\end{equation}
Here, $\sigma_i^- = |0\rangle_i\langle 1|$ and $\sigma_i^+ = |1\rangle_i\langle 0|$ are the lowering and raising operators for the $i$-th subsystem, respectively. The parameter $\Gamma_i > 0$ represents the decay rate characterizing the coupling of subsystem $i$ to its local zero-temperature reservoir. The physical mechanism described by this equation is straightforward: each two-level system undergoes independent radiative decay from its excited state $|1\rangle$ to its ground state $|0\rangle$ at a rate $\Gamma_i$, without additional pure dephasing channels beyond those intrinsic to the amplitude damping process. Since the subsystems do not interact directly, any dynamical correlations or collective behaviors arise solely from the quantum or classical correlations encoded in the initial state.

To investigate the influence of dissipation on the entanglement, we consider a class of initial states known as X-states. In the basis $\{|00\rangle, |01\rangle, |10\rangle, |11\rangle\}$, the density matrix $\rho_{AB}(0)$ takes the form
\begin{equation}
	\rho_{AB}(0) =\begin{pmatrix}
		&a(0) &0 &0 &w(0)\\
		&0 &b(0) &z(0) &0\\
		&0 &z(0)^* &c(0) &0\\
		&w(0)^* &0 &0 &d(0)
	\end{pmatrix},
\end{equation}
where the diagonal elements satisfy the normalization condition $a(t) + b(t) + c(t) + d(t)=1$. The requirements of positive semi-definiteness further impose the constraints $a(t)d(t) \ge |w(t)|^2$ and $b(t)c(t) \ge |z(t)|^2$. A crucial mathematical property of X-states is their form-invariance under many decoherence channels. For the specific dynamics of independent local amplitude damping at zero temperature, as governed by Eq.~(\ref{eq_lind}), this property ensures that an initially X-shaped state preserves its structure throughout the entire evolution. Furthermore, X-states have been extensively employed to explore various quantum correlation phenomena, such as ESD ~\cite{Bellomo:2007mms,Bellomo:2007euv,Pan:2025oqc,Cavalcante:2025wlm,Nakamura:2025mae,Rau:2009ufg,Yu:2007bwc}. To quantify the entanglement between subsystems A and B, we employ  the concurrence \(\mathcal{C}(\rho)\). For an X-state, the concurrence at time \(t\) is given by~\cite{Hill:1997pfa,Wootters:1997id,Yu:2007bwc}:
\begin{equation}
	\mathcal{C}(\rho(t)) = 2 \mathbf{max} \left\{ 0, |z(t)| - \sqrt{a(t)d(t)}, |w(t)| - \sqrt{b(t)c(t)}\right\} .
\end{equation}
In this study, we simplify the model by setting $z(0) = z(0)^* = 0$, which  implies  $z(t)=0$ for all $t$ (see Appendix). Under this condition, the expression for concurrence reduces to:
\begin{equation}
	\mathcal{C}(\rho(t))=2\mathbf{max}\left\{0,|w(t)|-\sqrt{b(t)c(t)}\right\}.
\end{equation}
The concurrence $\mathcal{C}(\rho(t))$ is therefore determined entirely by the off-diagonal element $w(t)$ and the product of diagonal elements $b(t)\cdot c(t)$. If we further impose the initial condition $b(0) = c(0) = 0$, the initial concurrence reduces to a function of $w(0)$ alone.
Consequently, \(w(0)\) assumes the role of an effective initial-temperature parameter, directly analogous to the initial temperature in the classical Mpemba effect. The remaining independent parameter \(d(0)\) then functions as a switch that controls whether this entanglement Mpemba effect (EME) occurs.

\emph{Results-}	{\it Symmetric dissipation with $\Gamma_A = \Gamma_B$}. We first consider the symmetric case where both subsystems share an identical decay rate, $\Gamma_A = \Gamma_B = \Gamma$. Under this assumption, the Lindblad master equation, Eq.~(\ref{eq_lind}), reduces to its simplest form. The ESD time, denoted as $\tau^*_{ESD}$, is defined by the condition $\mathcal{C}(\rho(\tau^*_{ESD})) = 0$. For the specific initial state considered ($z=0$), this leads to the relation
	\begin{equation}
		|w(\tau^*_{ESD})|=|w(0)|\gamma(\tau^*_{ESD})=d(0)\gamma(\tau^*_{ESD})(1-\gamma(\tau^*_{ESD})),
	\end{equation}
	where $\gamma(t) = e^{-\Gamma t}$. Solving this equation yields the exact ESD time
	\begin{equation}\label{eq_sysesd}
		\tau^*_{ESD}=-\frac{1}{\Gamma}ln\left(1-\frac{|w(0)|}{d(0)}\right).
	\end{equation}
We note that the positivity condition of the density matrix imposes the physical constraint $|w(0)| \le \sqrt{d(0)(1-d(0))}$. This constraint delineates two distinct dynamical regimes for the entanglement: when $|w(0)| \ge d(0)$, ESD  is entirely avoided, and the entanglement vanishes only asymptotically as $t \to \infty$. Conversely, when $|w(0)| < d(0)$, the entanglement drops abruptly to zero at a finite time $\tau^*_{\text{ESD}}$, even though the individual subsystems approach their respective ground states only asymptotically. Within this dissipative regime, the ESD time $\tau^*_{\text{ESD}}$ increases monotonically with the ratio $|w(0)|/d(0)$. In the critical boundary case where $|w(0)| = d(0)$, the sudden death time diverges ($\tau^*_{\text{ESD}} \to \infty$), marking the threshold where finite-time entanglement death is completely forestalled. Furthermore, consider two distinct initial states characterized by parameters $w_1(0),\, d_1(0)$ and $w_2(0),\, d_2(0)$, respectively. The crossing time $\tau_{\text{cross}}$, at which the concurrence trajectories of these two initial states coincide, i.e., $\mathcal{C}(\rho_1) = \mathcal{C}(\rho_2)$, can be determined analytically as
\begin{equation}\label{eq_syscross}
	\tau_{\text{cross}} = -\frac{1}{\Gamma}\ln\left(1 - \frac{|w_1(0)| - |w_2(0)|}{d_1(0) - d_2(0)}\right).
\end{equation}
A necessary condition for the two dynamical trajectories to cross during the evolution is that $\Delta w$ and $\Delta d$ have the same sign (both positive or both negative) and $|\Delta w| < |\Delta d|$, where $\Delta w = |w_1(0)| - |w_2(0)|$ and $\Delta d = d_1(0) - d_2(0)$.

\begin{figure}[t]
		\centering{\includegraphics[width=0.850\linewidth]{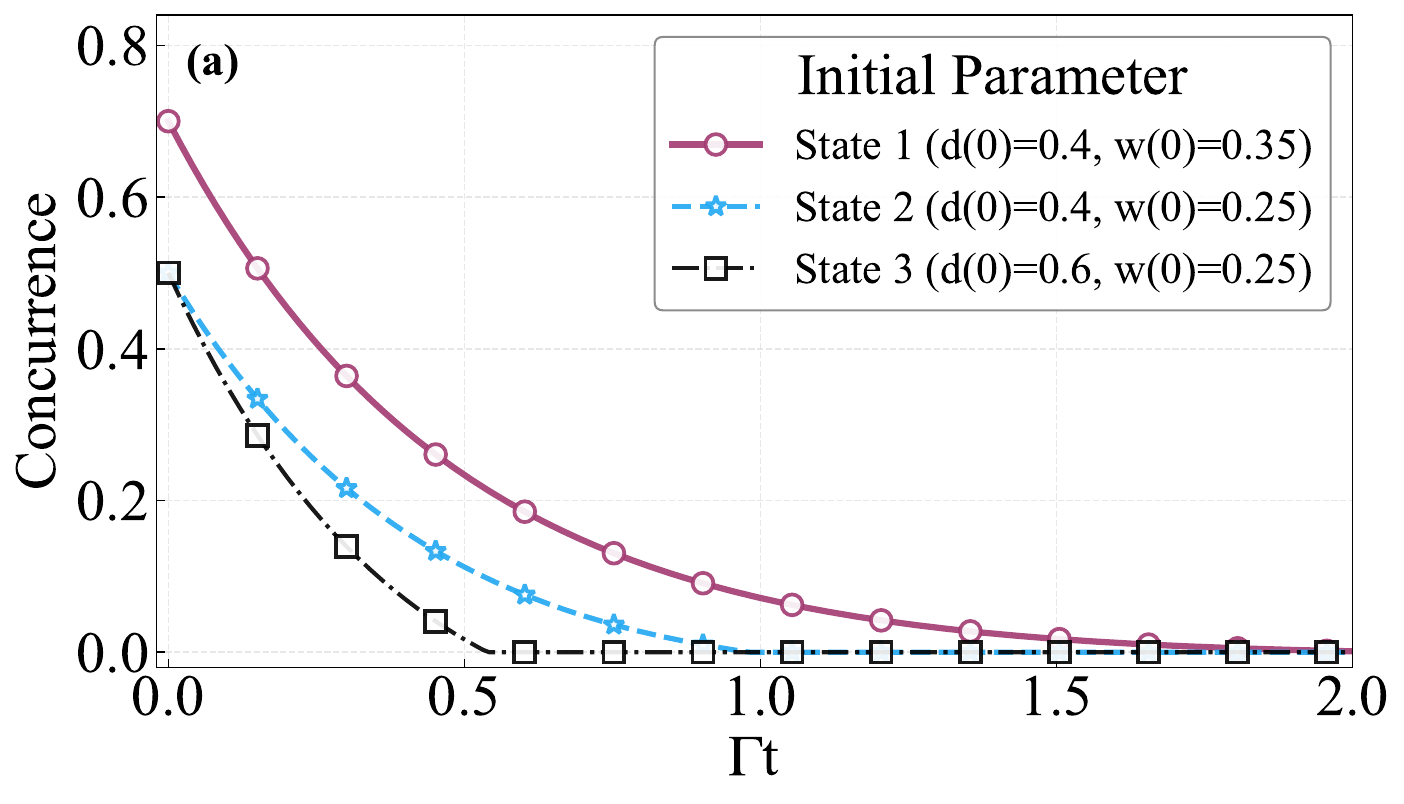}
			\includegraphics[width=0.850\linewidth]{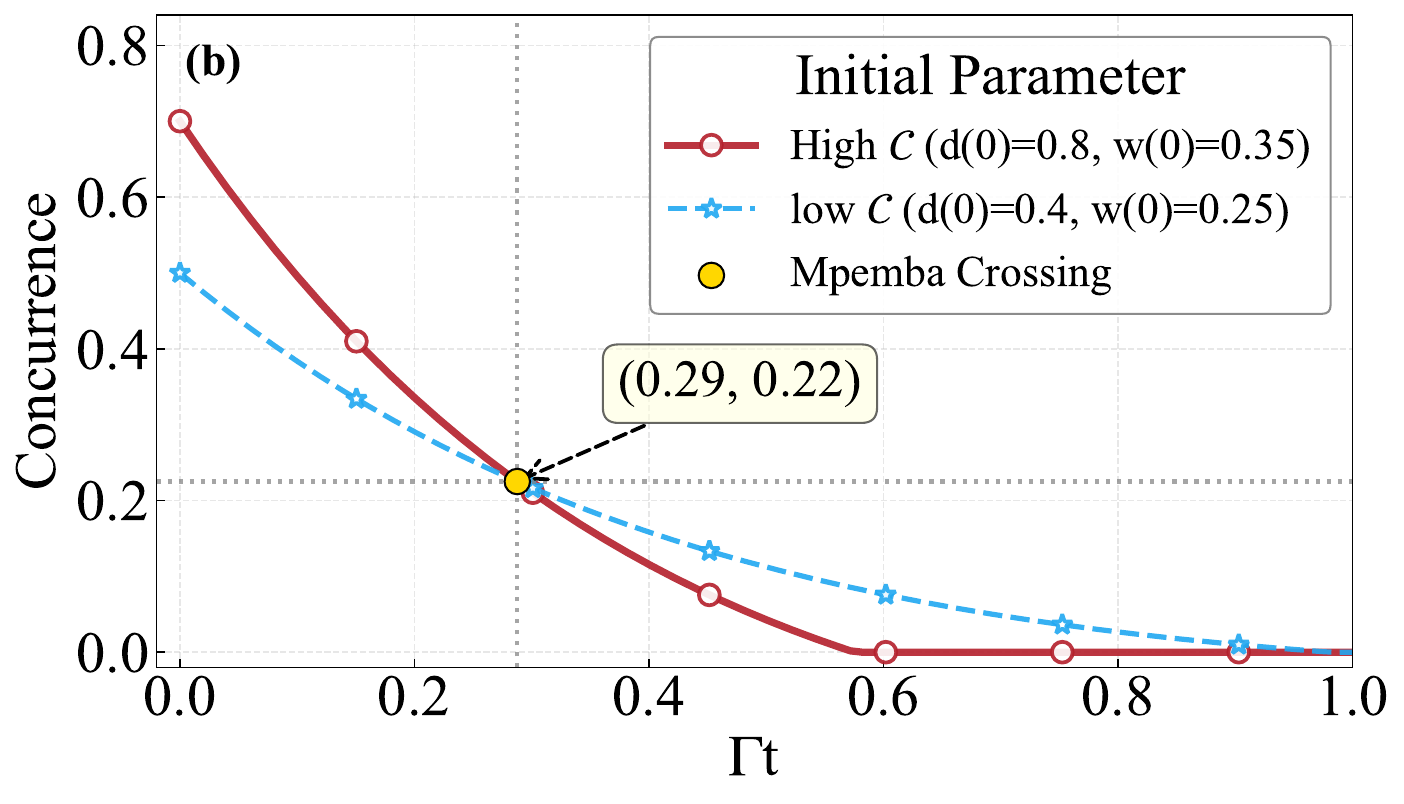}
			\caption{Decay of concurrence under symmetric dissipation $\Gamma_A = \Gamma_B = 1$. (a) Normal concurrence decay for three different initial states: $d(0)=0.4,\,0.4,\,0.6$ and $w(0)=0.35,\,0.25,\,0.25$, respectively. (b) Anomalous concurrence decay for two different initial states: $d_1(0)=0.4$, $w_1(0)=0.25$, and $d_2(0)=0.8$, $w_2(0)=0.35$. The yellow circle marks the crossing point at \((0.29, 0.22)\) (coordinates rounded to two decimal places).
			}\label{fig_EME}
		}
	\end{figure}

Fig.~\ref{fig_EME}(a) illustrates that the concurrence exhibits a monotonic decay without any crossing of trajectories during the evolution. As previously established, the initial concurrence is determined exclusively by the off-diagonal element $|w(0)|$. The blue and black curves share the same $|w(0)|$ but differ in $d(0)$. It is evident that a larger initial population in the excited state ($d(0)$) leads to a shorter ESD time $\tau^*_{\text{ESD}}$. This demonstrates that while $|w(0)|$ sets the initial concurrence, $d(0)$ governs the initial decay rate (or equivalently, the ESD time $\tau^*_{\text{ESD}}$).	We define the decay velocity of the concurrence as $v(t)$, which is given by the following relation
\begin{equation}
	v(t) = \frac{d\mathcal{C}(t)}{dt} = -2\Gamma \gamma \left[ |w(0)| - d(0) + 2\gamma d(0) \right].
\end{equation}
Subsequently, the initial decay velocity $v(0)$ depends on the sum of the parameters $|w(0)|$ and $d(0)$
\begin{equation}
	v(0) = -2\Gamma \left[ |w(0)| + d(0) \right].
\end{equation}
This result implies that for a fixed initial concurrence (which is determined solely by $|w(0)|$), the initial decay velocity can be tuned by varying the parameter $d(0)$. In the zero-temperature amplitude damping model, only qubits in the excited state can exchange energy with the environment by emitting photons. Consequently, a higher initial excited state population \(d(0)\) implies a larger number of available excitation quanta, which enhances the overall dissipative rate. This causes the system to transition more rapidly from the \(|11\rangle\) state to lower energy states, thereby accelerating the growth of the product term \(b(t)c(t)\). We now demonstrate that this feature can be exploited: by suitably tuning the system's initial parameters, a crossover phenomenon analogous to the Mpemba effect emerges in the evolution of the concurrence.

In Fig.~\ref{fig_EME}(b), we consider two initial states: a highly entangled state (red curve) with parameters $(d_2(0)=0.8, w_2(0)=0.35)$ and a state with lower initial concurrence (blue curve) with $(d_1(0)=0.4, w_1(0)=0.25)$. Their trajectories intersect at time $\tau_{\text{cross}}$, with the intersection point highlighted by a yellow circle. This phenomenon provides a quantum analogue of the Mpemba effect: under identical environmental conditions (fixed $\Gamma$), a system that is initially more highly entangled can reach the separable state faster than one with lower initial entanglement. The specific Mpemba crossing time $\tau_{\text{cross}}$ can be analytically determined using Eq.~(\ref{eq_syscross}).
	
{\it Asymmetric dissipation with $\Gamma_A \ne \Gamma_B$}.
Introducing asymmetric dissipation as a more general framework complicates the underlying master equation, yet it provides deeper insights into the dynamical characteristics of the EME. In the asymmetric regime ($\Gamma_A \neq \Gamma_B$), the concurrence between subsystems $A$ and $B$ can be expressed as (see Appendix)
\begin{equation}\label{eq_conc}
	\mathcal{C}(\rho(t)) = 2 \mathbf{max} \left\{ 0, \sqrt{\gamma_A \gamma_B}\left(|w(0)| - d(0)\sqrt{(1-\gamma_A)(1-\gamma_B)}\right)\right\},
\end{equation}
where $\gamma_i(t)=e^{-\Gamma_i t}$. 

\begin{figure}[t]
	\centering{\includegraphics[width=0.850\linewidth]{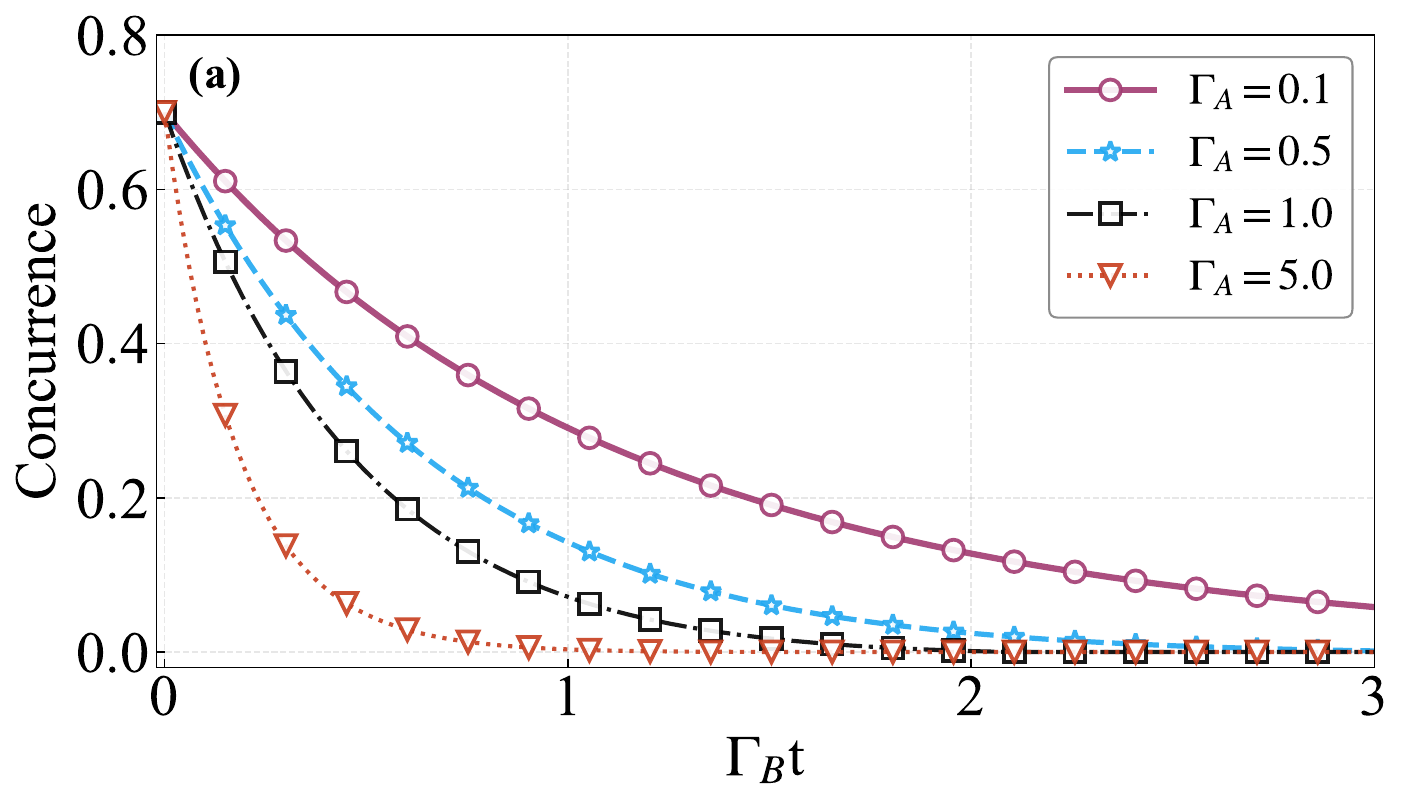}
		\includegraphics[width=0.850\linewidth]{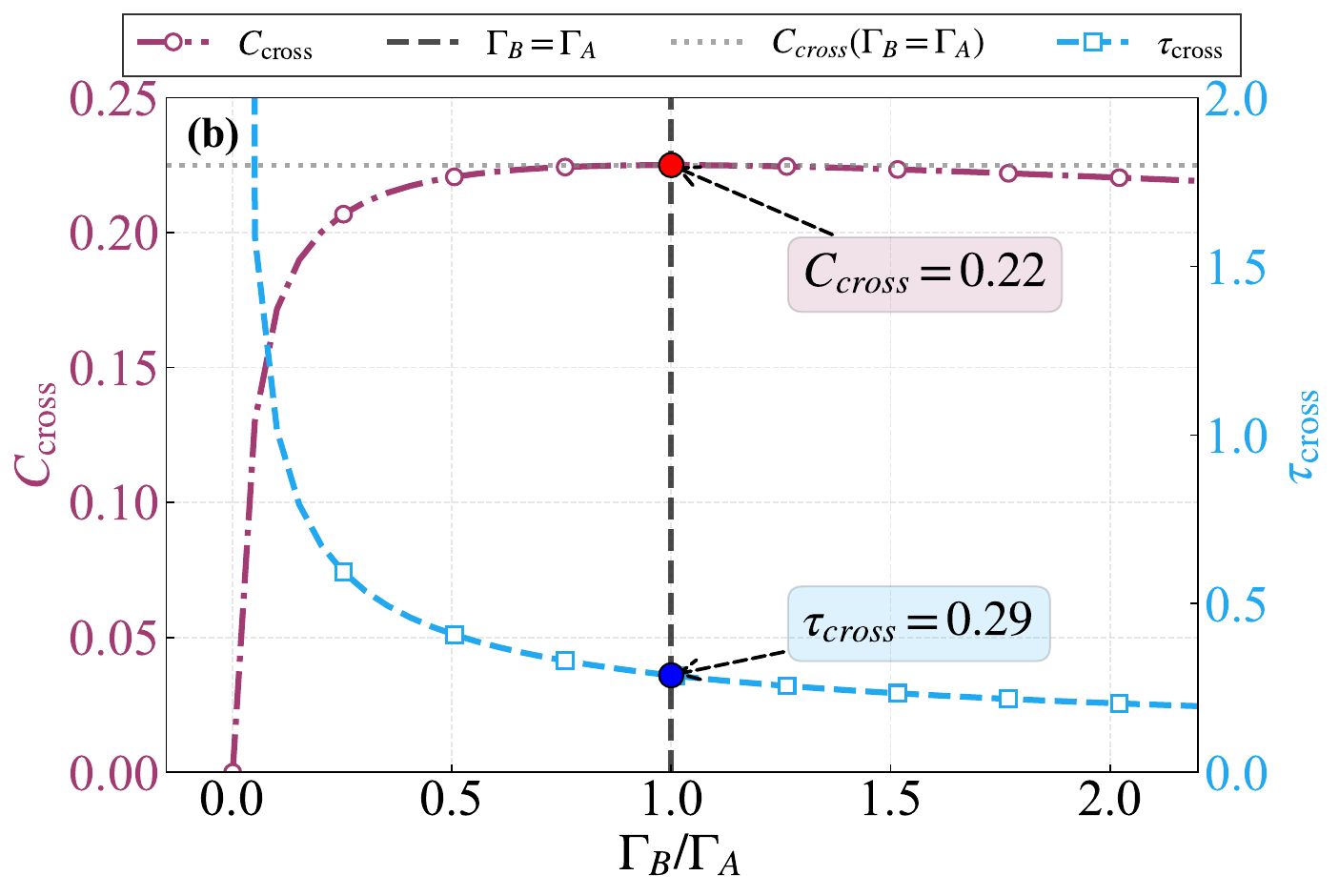}
		\caption{(a) Concurrence decay for the same initial state under four different values of \(\Gamma_A\): 0.1(pink circle), 0.5(blue star), 1(black square), 5(orange triangle), respectively. $w(0)=0.35,\,d(0)=0.4$. (b) The Mpemba crossing of concurrence $C_{\text{cross}}$ (pink circles) and the corresponding crossing time $\tau_{\text{cross}}$ (blue squares) vary with $\Gamma_B$. The initial state parameters are fixed as: $d_1(0)=0.4$, $w_1(0)=0.25$ and $d_2(0)=0.8$, $w_2(0)=0.35$.
		}\label{fig_AntiEME}
	}
\end{figure}
As illustrated in Fig.~\ref{fig_AntiEME}(a), for a fixed $\Gamma_B$, increasing $\Gamma_A$ accelerates the decay of concurrence toward sudden death, as a larger $\Gamma_A$ corresponds to enhanced environmental de-excitation. To systematically investigate the impact of this dissipation asymmetry on the EME, it is essential to determine the entanglement sudden death time $\tau^*_{\text{ESD}}$ and the concurrence trajectory crossing time $\tau_{\text{cross}}$. These timescales are uniquely defined by the conditions $C(\tau^*_{\text{ESD}}) = 0$ and $C(\rho_1(\tau_{\text{cross}})) = C(\rho_2(\tau_{\text{cross}}))$, which yields
\begin{align}
	&\frac{|w(0)|}{d(0)}-\sqrt{\left[1-\gamma_A(\tau^*_{\text{ESD}})\right]\left[1-\gamma_B(\tau^*_{\text{ESD}})\right]}=0, \label{eq_esd}\\
	&\frac{|w_1(0)|-|w_2(0)|}{d_1(0)-d_2(0)}-\sqrt{\left[1-\gamma_A(\tau_{\text{cross}})\right]\left[1-\gamma_B(\tau_{\text{cross}})\right]}=0.\label{eq_cro}
\end{align}
While these equations admit closed-form analytical solutions exclusively in the symmetric dissipation limit [Eqs.~(\ref{eq_sysesd}) and (\ref{eq_syscross})], they must be evaluated numerically in the presence of asymmetric dissipation. As shown in Fig.~\ref{fig_AntiEME}(b), we numerically evaluate the concurrence at the crossing point, $C_{\text{cross}}$ (pink dot-dashed line), and the corresponding crossing time, $\tau_{\text{cross}}$ (blue dashed line), as functions of the dissipation ratio $\Gamma_B/\Gamma_A$. As this ratio increases, the crossing time $\tau_{\text{cross}}$ decreases rapidly and monotonically, gradually leveling off after passing the symmetric dissipation threshold ($\Gamma_B/\Gamma_A = 1$). Furthermore, $C_{\text{cross}}$ exhibits a pronounced non-monotonic dependence on the ratio, reaching its maximum value exclusively at the symmetric setting ($\Gamma_B/\Gamma_A = 1$). This indicates that the symmetric configuration not only maximizes the entanglement preserved at the crossing point but also significantly accelerates the occurrence of the crossing behavior.

\emph{Discussion-}In classical thermodynamics, the Mpemba effect manifests as a counterintuitive phenomenon where a hotter system cools or freezes faster than a colder counterpart. Frequently classified as an anomalous relaxation phenomenon~\cite{PhysRevX.9.021060,Medina:2024yep,doi:10.1073/pnas.2118484119}, it occurs exclusively within a restricted regime of initial temperatures, whereas conventional, intuitive relaxation dominates the more general parameter space. A striking parallel emerges in the quantum domain for the EME investigated here. 

\begin{figure}[t]
	\centering{\includegraphics[width=0.86\linewidth]{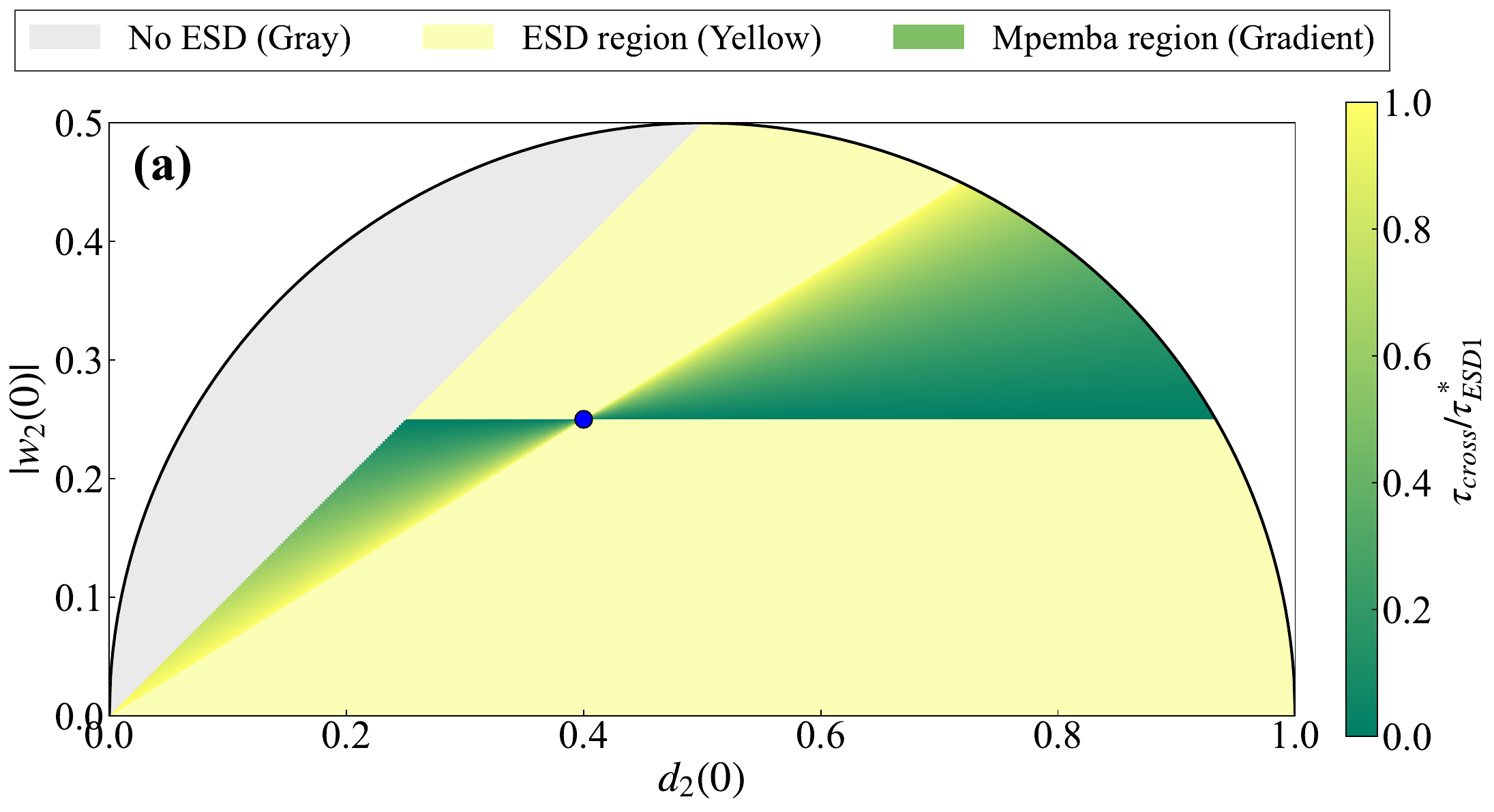}
		\includegraphics[width=0.86\linewidth]{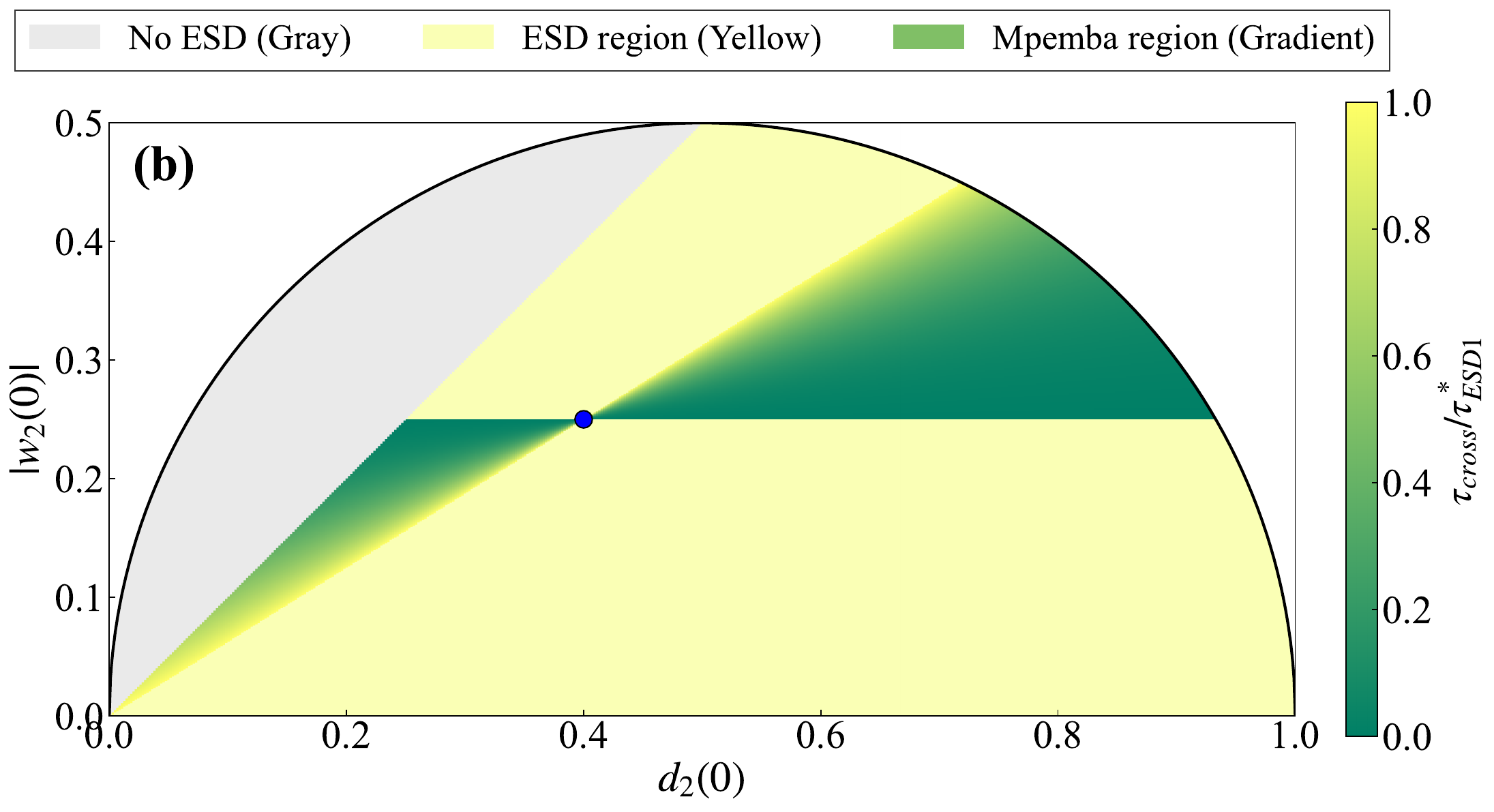}
		\caption{Phase diagrams: solid black semicircle is physical boundary (positivity condition). Grey area: no ESD. Green gradient: quantum Mpemba crossing. Yellow: ESD without Mpemba effect. Blue dot marks reference parameters ($|w_1(0)| = 0.25$, $d_1(0) = 0.4$). Panels: (a) symmetric dissipation ($\Gamma_A = \Gamma_B = 1$), (b) asymmetric dissipation ($\Gamma_A = 0.1$, $\Gamma_B = 10$).}\label{fig_phase}
	}
\end{figure}

As illustrated in the phase diagrams of Figs.~\ref{fig_phase}(a) and (b), by fixing the reference state configuration at $|w_1(0)| = 0.25$ and $d_1(0) = 0.4$, we map the onset of the EME in the parameter space spanned by the initial state variables $|w_2(0)|$ and $d_2(0)$ under both symmetric and asymmetric dissipation. Notably, the EME is confined to a sharply bounded, limited region, conceptually mirroring the `anomalous' nature of the classical thermodynamic effect. Furthermore, our results reveal that environmental asymmetry merely modulates the characteristic timescale over which the EME unfolds, leaving the phase boundaries of the permissible parameter regime entirely invariant.

\emph{Conclusion-}In summary, we have uncovered a distinct quantum Mpemba effect in  a dissipative quantum system. Under local amplitude damping, a more strongly entangled initial state can collapse into a completely separable state faster than a weakly entangled one. Phase diagram analysis reveals that the Mpemba effect manifests exclusively within a restricted regime of physically accessible parameters, closely echoing the classical thermodynamic Mpemba effect, which similarly occurs only within specific initial temperature intervals. Crucially, while the environmental asymmetry between the two subsystems modulates the timescales of the decay, it leaves the phase boundaries of the Mpemba regime invariant. 
We show that tuning the initial balance between coherence and excited-state population can regulate the lifetime of quantum resources in dissipative environments, with direct implications for quantum technologies from tabletop to space. These findings motivate experimental tests of entanglement-lifetime inversion and further investigation of its robustness under more general noise and coupling conditions.
 \\
 
\emph{Acknowledgements-}
    This work was supported by the National Natural Science Foundation of China under Grants No. 12475051, No. 12421005 and No. 12575050; the science and technology innovation Program of Hunan Province under grant No. 2024RC1050;  the innovative research group of Hunan Province under Grant No. 2024JJ1006; the Hangzhou Leading Youth Innovation and Entrepreneurship Team project  under Grant No. TD2024005; and the scientific research start-up funds of Hangzhou Normal University: 4245C50224204016.


\begin{thebibliography}{61}%
	\makeatletter
	\providecommand \@ifxundefined [1]{%
		\@ifx{#1\undefined}
	}%
	\providecommand \@ifnum [1]{%
		\ifnum #1\expandafter \@firstoftwo
		\else \expandafter \@secondoftwo
		\fi
	}%
	\providecommand \@ifx [1]{%
		\ifx #1\expandafter \@firstoftwo
		\else \expandafter \@secondoftwo
		\fi
	}%
	\providecommand \natexlab [1]{#1}%
	\providecommand \enquote  [1]{``#1''}%
	\providecommand \bibnamefont  [1]{#1}%
	\providecommand \bibfnamefont [1]{#1}%
	\providecommand \citenamefont [1]{#1}%
	\providecommand \href@noop [0]{\@secondoftwo}%
	\providecommand \href [0]{\begingroup \@sanitize@url \@href}%
	\providecommand \@href[1]{\@@startlink{#1}\@@href}%
	\providecommand \@@href[1]{\endgroup#1\@@endlink}%
	\providecommand \@sanitize@url [0]{\catcode `\\12\catcode `\$12\catcode
		`\&12\catcode `\#12\catcode `\^12\catcode `\_12\catcode `\%12\relax}%
	\providecommand \@@startlink[1]{}%
	\providecommand \@@endlink[0]{}%
	\providecommand \url  [0]{\begingroup\@sanitize@url \@url }%
	\providecommand \@url [1]{\endgroup\@href {#1}{\urlprefix }}%
	\providecommand \urlprefix  [0]{URL }%
	\providecommand \Eprint [0]{\href }%
	\providecommand \doibase [0]{https://doi.org/}%
	\providecommand \selectlanguage [0]{\@gobble}%
	\providecommand \bibinfo  [0]{\@secondoftwo}%
	\providecommand \bibfield  [0]{\@secondoftwo}%
	\providecommand \translation [1]{[#1]}%
	\providecommand \BibitemOpen [0]{}%
	\providecommand \bibitemStop [0]{}%
	\providecommand \bibitemNoStop [0]{.\EOS\space}%
	\providecommand \EOS [0]{\spacefactor3000\relax}%
	\providecommand \BibitemShut  [1]{\csname bibitem#1\endcsname}%
	\let\auto@bib@innerbib\@empty
	\bibitem [{\citenamefont {Mpemba}\ and\ \citenamefont
		{Osborne}(1969)}]{EBMpemba:1969}%
	\BibitemOpen
	\bibfield  {author} {\bibinfo {author} {\bibfnamefont {E.~B.}\ \bibnamefont
			{Mpemba}}\ and\ \bibinfo {author} {\bibfnamefont {D.~G.}\ \bibnamefont
			{Osborne}},\ }\bibfield  {title} {\bibinfo {title} {Cool?},\ }\href
	{https://doi.org/10.1088/0031-9120/4/3/312} {\bibfield  {journal} {\bibinfo
			{journal} {Phys. Educ.}\ }\textbf {\bibinfo {volume} {4}},\ \bibinfo {pages}
		{172} (\bibinfo {year} {1969})}\BibitemShut {NoStop}%
	\bibitem [{\citenamefont {Lu}\ and\ \citenamefont
		{Raz}(2017)}]{doi:10.1073/pnas.1701264114}%
	\BibitemOpen
	\bibfield  {author} {\bibinfo {author} {\bibfnamefont {Z.}~\bibnamefont
			{Lu}}\ and\ \bibinfo {author} {\bibfnamefont {O.}~\bibnamefont {Raz}},\
	}\bibfield  {title} {\bibinfo {title} {Nonequilibrium thermodynamics of the
			markovian mpemba effect and its inverse},\ }\href
	{https://doi.org/10.1073/pnas.1701264114} {\bibfield  {journal} {\bibinfo
			{journal} {Proc. Natl. Acad. Sci. U.S.A.}\ }\textbf {\bibinfo {volume}
			{114}},\ \bibinfo {pages} {5083} (\bibinfo {year} {2017})}\BibitemShut
	{NoStop}%
	\bibitem [{\citenamefont {Beato}\ and\ \citenamefont
		{Teza}(2026)}]{Beato:2025dbm}%
	\BibitemOpen
	\bibfield  {author} {\bibinfo {author} {\bibfnamefont {N.}~\bibnamefont
			{Beato}}\ and\ \bibinfo {author} {\bibfnamefont {G.}~\bibnamefont {Teza}},\
	}\bibfield  {title} {\bibinfo {title} {{Relaxation Control of Open Quantum
				Systems}},\ }\href {https://doi.org/10.1103/4frd-ck2z} {\bibfield  {journal}
		{\bibinfo  {journal} {Phys. Rev. Lett.}\ }\textbf {\bibinfo {volume} {136}},\
		\bibinfo {pages} {070401} (\bibinfo {year} {2026})},\ \Eprint
	{https://arxiv.org/abs/2507.15948} {arXiv:2507.15948 [quant-ph]} \BibitemShut
	{NoStop}%
	\bibitem [{\citenamefont {Bao}(2026)}]{Bao:2025ocj}%
	\BibitemOpen
	\bibfield  {author} {\bibinfo {author} {\bibfnamefont {R.}~\bibnamefont
			{Bao}},\ }\bibfield  {title} {\bibinfo {title} {{Initial-State Typicality in
				Quantum Relaxation}},\ }\href {https://doi.org/10.1103/wgr5-lb6b} {\bibfield
		{journal} {\bibinfo  {journal} {Phys. Rev. Lett.}\ }\textbf {\bibinfo
			{volume} {136}},\ \bibinfo {pages} {070402} (\bibinfo {year} {2026})},\
	\Eprint {https://arxiv.org/abs/2511.01709} {arXiv:2511.01709 [quant-ph]}
	\BibitemShut {NoStop}%
	\bibitem [{\citenamefont {Nava}\ and\ \citenamefont
		{Egger}(2025)}]{Nava:2025fdd}%
	\BibitemOpen
	\bibfield  {author} {\bibinfo {author} {\bibfnamefont {A.}~\bibnamefont
			{Nava}}\ and\ \bibinfo {author} {\bibfnamefont {R.}~\bibnamefont {Egger}},\
	}\bibfield  {title} {\bibinfo {title} {{Pontus-Mpemba Effects}},\ }\href
	{https://doi.org/10.1103/hhgj-89gj} {\bibfield  {journal} {\bibinfo
			{journal} {Phys. Rev. Lett.}\ }\textbf {\bibinfo {volume} {135}},\ \bibinfo
		{pages} {140404} (\bibinfo {year} {2025})},\ \Eprint
	{https://arxiv.org/abs/2505.14622} {arXiv:2505.14622 [quant-ph]} \BibitemShut
	{NoStop}%
	\bibitem [{\citenamefont {Carollo}\ \emph {et~al.}(2021)\citenamefont
		{Carollo}, \citenamefont {Lasanta},\ and\ \citenamefont
		{Lesanovsky}}]{Carollo:2021hew}%
	\BibitemOpen
	\bibfield  {author} {\bibinfo {author} {\bibfnamefont {F.}~\bibnamefont
			{Carollo}}, \bibinfo {author} {\bibfnamefont {A.}~\bibnamefont {Lasanta}},\
		and\ \bibinfo {author} {\bibfnamefont {I.}~\bibnamefont {Lesanovsky}},\
	}\bibfield  {title} {\bibinfo {title} {{Exponentially accelerated approach to
				stationarity in Markovian open quantum systems through the Mpemba effect}},\
	}\href {https://doi.org/10.1103/PhysRevLett.127.060401} {\bibfield  {journal}
		{\bibinfo  {journal} {Phys. Rev. Lett.}\ }\textbf {\bibinfo {volume} {127}},\
		\bibinfo {pages} {060401} (\bibinfo {year} {2021})},\ \Eprint
	{https://arxiv.org/abs/2103.05020} {arXiv:2103.05020 [quant-ph]} \BibitemShut
	{NoStop}%
	\bibitem [{\citenamefont {Longhi}(2024)}]{Longhi:2024omn}%
	\BibitemOpen
	\bibfield  {author} {\bibinfo {author} {\bibfnamefont {S.}~\bibnamefont
			{Longhi}},\ }\bibfield  {title} {\bibinfo {title} {{Photonic Mpemba
				effect}},\ }\href {https://doi.org/10.1364/OL.532503} {\bibfield  {journal}
		{\bibinfo  {journal} {Opt. Lett.}\ }\textbf {\bibinfo {volume} {49}},\
		\bibinfo {pages} {5188} (\bibinfo {year} {2024})},\ \Eprint
	{https://arxiv.org/abs/2408.03296} {arXiv:2408.03296 [physics.optics]}
	\BibitemShut {NoStop}%
	\bibitem [{\citenamefont {Van~Vu}\ and\ \citenamefont
		{Hayakawa}(2025)}]{VanVu:2024zps}%
	\BibitemOpen
	\bibfield  {author} {\bibinfo {author} {\bibfnamefont {T.}~\bibnamefont
			{Van~Vu}}\ and\ \bibinfo {author} {\bibfnamefont {H.}~\bibnamefont
			{Hayakawa}},\ }\bibfield  {title} {\bibinfo {title} {{Thermomajorization
				Mpemba Effect}},\ }\href {https://doi.org/10.1103/PhysRevLett.134.107101}
	{\bibfield  {journal} {\bibinfo  {journal} {Phys. Rev. Lett.}\ }\textbf
		{\bibinfo {volume} {134}},\ \bibinfo {pages} {107101} (\bibinfo {year}
		{2025})},\ \Eprint {https://arxiv.org/abs/2410.06686} {arXiv:2410.06686
		[cond-mat.stat-mech]} \BibitemShut {NoStop}%
	\bibitem [{\citenamefont {Bao}\ and\ \citenamefont {Hou}(2025)}]{Bao:2022ffs}%
	\BibitemOpen
	\bibfield  {author} {\bibinfo {author} {\bibfnamefont {R.}~\bibnamefont
			{Bao}}\ and\ \bibinfo {author} {\bibfnamefont {Z.}~\bibnamefont {Hou}},\
	}\bibfield  {title} {\bibinfo {title} {{Accelerating Quantum Relaxation via
				Temporary Reset: A Mpemba-Inspired Approach}},\ }\href
	{https://doi.org/10.1103/g94p-7421} {\bibfield  {journal} {\bibinfo
			{journal} {Phys. Rev. Lett.}\ }\textbf {\bibinfo {volume} {135}},\ \bibinfo
		{pages} {150403} (\bibinfo {year} {2025})},\ \Eprint
	{https://arxiv.org/abs/2212.11170} {arXiv:2212.11170 [quant-ph]} \BibitemShut
	{NoStop}%
	\bibitem [{\citenamefont {Wang}\ and\ \citenamefont
		{Wang}(2024)}]{Wang:2024jlr}%
	\BibitemOpen
	\bibfield  {author} {\bibinfo {author} {\bibfnamefont {X.}~\bibnamefont
			{Wang}}\ and\ \bibinfo {author} {\bibfnamefont {J.}~\bibnamefont {Wang}},\
	}\bibfield  {title} {\bibinfo {title} {{Mpemba effects in nonequilibrium open
				quantum systems}},\ }\href {https://doi.org/10.1103/PhysRevResearch.6.033330}
	{\bibfield  {journal} {\bibinfo  {journal} {Phys. Rev. Res.}\ }\textbf
		{\bibinfo {volume} {6}},\ \bibinfo {pages} {033330} (\bibinfo {year}
		{2024})},\ \Eprint {https://arxiv.org/abs/2401.14259} {arXiv:2401.14259
		[quant-ph]} \BibitemShut {NoStop}%
	\bibitem [{\citenamefont {Strachan}\ \emph {et~al.}(2025)\citenamefont
		{Strachan}, \citenamefont {Purkayastha},\ and\ \citenamefont
		{Clark}}]{Strachan:2024mva}%
	\BibitemOpen
	\bibfield  {author} {\bibinfo {author} {\bibfnamefont {D.~J.}\ \bibnamefont
			{Strachan}}, \bibinfo {author} {\bibfnamefont {A.}~\bibnamefont
			{Purkayastha}},\ and\ \bibinfo {author} {\bibfnamefont {S.~R.}\ \bibnamefont
			{Clark}},\ }\bibfield  {title} {\bibinfo {title} {{Non-Markovian Quantum
				Mpemba Effect}},\ }\href {https://doi.org/10.1103/PhysRevLett.134.220403}
	{\bibfield  {journal} {\bibinfo  {journal} {Phys. Rev. Lett.}\ }\textbf
		{\bibinfo {volume} {134}},\ \bibinfo {pages} {220403} (\bibinfo {year}
		{2025})},\ \Eprint {https://arxiv.org/abs/2402.05756} {arXiv:2402.05756
		[quant-ph]} \BibitemShut {NoStop}%
	\bibitem [{\citenamefont {Moroder}\ \emph {et~al.}(2024)\citenamefont
		{Moroder}, \citenamefont {Culhane}, \citenamefont {Zawadzki},\ and\
		\citenamefont {Goold}}]{Moroder:2024jjh}%
	\BibitemOpen
	\bibfield  {author} {\bibinfo {author} {\bibfnamefont {M.}~\bibnamefont
			{Moroder}}, \bibinfo {author} {\bibfnamefont {O.}~\bibnamefont {Culhane}},
		\bibinfo {author} {\bibfnamefont {K.}~\bibnamefont {Zawadzki}},\ and\
		\bibinfo {author} {\bibfnamefont {J.}~\bibnamefont {Goold}},\ }\bibfield
	{title} {\bibinfo {title} {{Thermodynamics of the Quantum Mpemba Effect}},\
	}\href {https://doi.org/10.1103/PhysRevLett.133.140404} {\bibfield  {journal}
		{\bibinfo  {journal} {Phys. Rev. Lett.}\ }\textbf {\bibinfo {volume} {133}},\
		\bibinfo {pages} {140404} (\bibinfo {year} {2024})},\ \Eprint
	{https://arxiv.org/abs/2403.16959} {arXiv:2403.16959 [quant-ph]} \BibitemShut
	{NoStop}%
	\bibitem [{\citenamefont {Chatterjee}\ \emph {et~al.}(2023)\citenamefont
		{Chatterjee}, \citenamefont {Takada},\ and\ \citenamefont
		{Hayakawa}}]{Chatterjee:2023lgx}%
	\BibitemOpen
	\bibfield  {author} {\bibinfo {author} {\bibfnamefont {A.~K.}\ \bibnamefont
			{Chatterjee}}, \bibinfo {author} {\bibfnamefont {S.}~\bibnamefont {Takada}},\
		and\ \bibinfo {author} {\bibfnamefont {H.}~\bibnamefont {Hayakawa}},\
	}\bibfield  {title} {\bibinfo {title} {{Quantum Mpemba Effect in a Quantum
				Dot with Reservoirs}},\ }\href
	{https://doi.org/10.1103/PhysRevLett.131.080402} {\bibfield  {journal}
		{\bibinfo  {journal} {Phys. Rev. Lett.}\ }\textbf {\bibinfo {volume} {131}},\
		\bibinfo {pages} {080402} (\bibinfo {year} {2023})},\ \Eprint
	{https://arxiv.org/abs/2304.02411} {arXiv:2304.02411 [cond-mat.stat-mech]}
	\BibitemShut {NoStop}%
	\bibitem [{\citenamefont {Chattopadhyay}\ \emph {et~al.}(2026)\citenamefont
		{Chattopadhyay}, \citenamefont {Santos},\ and\ \citenamefont
		{Misra}}]{Chattopadhyay:2026qgl}%
	\BibitemOpen
	\bibfield  {author} {\bibinfo {author} {\bibfnamefont {P.}~\bibnamefont
			{Chattopadhyay}}, \bibinfo {author} {\bibfnamefont {J.~F.~G.}\ \bibnamefont
			{Santos}},\ and\ \bibinfo {author} {\bibfnamefont {A.}~\bibnamefont
			{Misra}},\ }\bibfield  {title} {\bibinfo {title} {{Anomaly to Resource: The
				Mpemba Effect in Quantum Thermometry}},\ }\href@noop {} {\  (\bibinfo {year}
		{2026})},\ \Eprint {https://arxiv.org/abs/2601.05046} {arXiv:2601.05046
		[quant-ph]} \BibitemShut {NoStop}%
	\bibitem [{\citenamefont {Torrente}\ \emph {et~al.}(2019)\citenamefont
		{Torrente}, \citenamefont {L\'opez-Casta\~no}, \citenamefont {Lasanta},
		\citenamefont {Reyes}, \citenamefont {Prados},\ and\ \citenamefont
		{Santos}}]{PhysRevE.99.060901}%
	\BibitemOpen
	\bibfield  {author} {\bibinfo {author} {\bibfnamefont {A.}~\bibnamefont
			{Torrente}}, \bibinfo {author} {\bibfnamefont {M.~A.}\ \bibnamefont
			{L\'opez-Casta\~no}}, \bibinfo {author} {\bibfnamefont {A.}~\bibnamefont
			{Lasanta}}, \bibinfo {author} {\bibfnamefont {F.~V.}\ \bibnamefont {Reyes}},
		\bibinfo {author} {\bibfnamefont {A.}~\bibnamefont {Prados}},\ and\ \bibinfo
		{author} {\bibfnamefont {A.}~\bibnamefont {Santos}},\ }\bibfield  {title}
	{\bibinfo {title} {Large mpemba-like effect in a gas of inelastic rough hard
			spheres},\ }\href {https://doi.org/10.1103/PhysRevE.99.060901} {\bibfield
		{journal} {\bibinfo  {journal} {Phys. Rev. E}\ }\textbf {\bibinfo {volume}
			{99}},\ \bibinfo {pages} {060901(R)} (\bibinfo {year} {2019})}\BibitemShut
	{NoStop}%
	\bibitem [{\citenamefont {Kumar}\ and\ \citenamefont
		{Bechhoefer}(2020)}]{Kumar2020ExponentiallyFC}%
	\BibitemOpen
	\bibfield  {author} {\bibinfo {author} {\bibfnamefont {A.}~\bibnamefont
			{Kumar}}\ and\ \bibinfo {author} {\bibfnamefont {J.}~\bibnamefont
			{Bechhoefer}},\ }\bibfield  {title} {\bibinfo {title} {Exponentially faster
			cooling in a colloidal system},\ }\href
	{https://api.semanticscholar.org/CorpusID:221006445} {\bibfield  {journal}
		{\bibinfo  {journal} {Nature}\ }\textbf {\bibinfo {volume} {584}},\ \bibinfo
		{pages} {64 } (\bibinfo {year} {2020})}\BibitemShut {NoStop}%
	\bibitem [{\citenamefont {Aharony~Shapira}\ \emph {et~al.}(2024)\citenamefont
		{Aharony~Shapira}, \citenamefont {Shapira}, \citenamefont {Markov},
		\citenamefont {Teza}, \citenamefont {Akerman}, \citenamefont {Raz},\ and\
		\citenamefont {Ozeri}}]{AharonyShapira:2024nrt}%
	\BibitemOpen
	\bibfield  {author} {\bibinfo {author} {\bibfnamefont {S.}~\bibnamefont
			{Aharony~Shapira}}, \bibinfo {author} {\bibfnamefont {Y.}~\bibnamefont
			{Shapira}}, \bibinfo {author} {\bibfnamefont {J.}~\bibnamefont {Markov}},
		\bibinfo {author} {\bibfnamefont {G.}~\bibnamefont {Teza}}, \bibinfo {author}
		{\bibfnamefont {N.}~\bibnamefont {Akerman}}, \bibinfo {author} {\bibfnamefont
			{O.}~\bibnamefont {Raz}},\ and\ \bibinfo {author} {\bibfnamefont
			{R.}~\bibnamefont {Ozeri}},\ }\bibfield  {title} {\bibinfo {title} {{Inverse
				Mpemba Effect Demonstrated on a Single Trapped Ion Qubit}},\ }\href
	{https://doi.org/10.1103/PhysRevLett.133.010403} {\bibfield  {journal}
		{\bibinfo  {journal} {Phys. Rev. Lett.}\ }\textbf {\bibinfo {volume} {133}},\
		\bibinfo {pages} {010403} (\bibinfo {year} {2024})},\ \Eprint
	{https://arxiv.org/abs/2401.05830} {arXiv:2401.05830 [quant-ph]} \BibitemShut
	{NoStop}%
	\bibitem [{\citenamefont {Zhang}\ \emph {et~al.}(2025)\citenamefont {Zhang}
		\emph {et~al.}}]{Zhang:2024juc}%
	\BibitemOpen
	\bibfield  {author} {\bibinfo {author} {\bibfnamefont {J.}~\bibnamefont
			{Zhang}} \emph {et~al.},\ }\bibfield  {title} {\bibinfo {title} {{Observation
				of quantum strong Mpemba effect}},\ }\href
	{https://doi.org/10.1038/s41467-024-54303-0} {\bibfield  {journal} {\bibinfo
			{journal} {Nature Commun.}\ }\textbf {\bibinfo {volume} {16}},\ \bibinfo
		{pages} {301} (\bibinfo {year} {2025})},\ \Eprint
	{https://arxiv.org/abs/2401.15951} {arXiv:2401.15951 [quant-ph]} \BibitemShut
	{NoStop}%
	\bibitem [{\citenamefont {Ares}\ \emph {et~al.}(2025)\citenamefont {Ares},
		\citenamefont {Calabrese},\ and\ \citenamefont {Murciano}}]{Ares:2025onj}%
	\BibitemOpen
	\bibfield  {author} {\bibinfo {author} {\bibfnamefont {F.}~\bibnamefont
			{Ares}}, \bibinfo {author} {\bibfnamefont {P.}~\bibnamefont {Calabrese}},\
		and\ \bibinfo {author} {\bibfnamefont {S.}~\bibnamefont {Murciano}},\
	}\bibfield  {title} {\bibinfo {title} {{The quantum Mpemba effects}},\ }\href
	{https://doi.org/10.1038/s42254-025-00838-0} {\bibfield  {journal} {\bibinfo
			{journal} {Nature Rev. Phys.}\ }\textbf {\bibinfo {volume} {7}},\ \bibinfo
		{pages} {451} (\bibinfo {year} {2025})},\ \Eprint
	{https://arxiv.org/abs/2502.08087} {arXiv:2502.08087 [cond-mat.stat-mech]}
	\BibitemShut {NoStop}%
	\bibitem [{\citenamefont {Teza}\ \emph {et~al.}(2026)\citenamefont {Teza},
		\citenamefont {Bechhoefer}, \citenamefont {Lasanta}, \citenamefont {Raz},\
		and\ \citenamefont {Vucelja}}]{Teza:2025azr}%
	\BibitemOpen
	\bibfield  {author} {\bibinfo {author} {\bibfnamefont {G.}~\bibnamefont
			{Teza}}, \bibinfo {author} {\bibfnamefont {J.}~\bibnamefont {Bechhoefer}},
		\bibinfo {author} {\bibfnamefont {A.}~\bibnamefont {Lasanta}}, \bibinfo
		{author} {\bibfnamefont {O.}~\bibnamefont {Raz}},\ and\ \bibinfo {author}
		{\bibfnamefont {M.}~\bibnamefont {Vucelja}},\ }\bibfield  {title} {\bibinfo
		{title} {{Speedups in nonequilibrium thermal relaxation: Mpemba and related
				effects}},\ }\href {https://doi.org/10.1016/j.physrep.2025.10.009} {\bibfield
		{journal} {\bibinfo  {journal} {Phys. Rept.}\ }\textbf {\bibinfo {volume}
			{1164}},\ \bibinfo {pages} {1} (\bibinfo {year} {2026})},\ \Eprint
	{https://arxiv.org/abs/2502.01758} {arXiv:2502.01758 [cond-mat.stat-mech]}
	\BibitemShut {NoStop}%
	\bibitem [{\citenamefont {Bechhoefer}\ \emph {et~al.}(2021)\citenamefont
		{Bechhoefer}, \citenamefont {Kumar},\ and\ \citenamefont
		{Ch{\'e}trite}}]{bechhoefer2021fresh}%
	\BibitemOpen
	\bibfield  {author} {\bibinfo {author} {\bibfnamefont {J.}~\bibnamefont
			{Bechhoefer}}, \bibinfo {author} {\bibfnamefont {A.}~\bibnamefont {Kumar}},\
		and\ \bibinfo {author} {\bibfnamefont {R.}~\bibnamefont {Ch{\'e}trite}},\
	}\bibfield  {title} {\bibinfo {title} {A fresh understanding of the mpemba
			effect},\ }\href {https://doi.org/10.1038/s42254-021-00349-8} {\bibfield
		{journal} {\bibinfo  {journal} {Nat. Rev. Phys.}\ }\textbf {\bibinfo {volume}
			{3}},\ \bibinfo {pages} {534} (\bibinfo {year} {2021})}\BibitemShut {NoStop}%
	\bibitem [{\citenamefont {Yu}\ \emph {et~al.}(2025{\natexlab{a}})\citenamefont
		{Yu}, \citenamefont {Liu},\ and\ \citenamefont {Zhang}}]{Yu:2025vth}%
	\BibitemOpen
	\bibfield  {author} {\bibinfo {author} {\bibfnamefont {H.}~\bibnamefont
			{Yu}}, \bibinfo {author} {\bibfnamefont {S.}~\bibnamefont {Liu}},\ and\
		\bibinfo {author} {\bibfnamefont {S.-X.}\ \bibnamefont {Zhang}},\ }\bibfield
	{title} {\bibinfo {title} {{Quantum Mpemba effects from symmetry
				perspectives}},\ }\href {https://doi.org/10.1007/s43673-025-00157-7}
	{\bibfield  {journal} {\bibinfo  {journal} {AAPPS Bull.}\ }\textbf {\bibinfo
			{volume} {35}},\ \bibinfo {pages} {17} (\bibinfo {year}
		{2025}{\natexlab{a}})},\ \Eprint {https://arxiv.org/abs/2507.02301}
	{arXiv:2507.02301 [quant-ph]} \BibitemShut {NoStop}%
	\bibitem [{\citenamefont {Chang}\ \emph {et~al.}(2026)\citenamefont {Chang},
		\citenamefont {Yin}, \citenamefont {Zhang},\ and\ \citenamefont
		{Li}}]{Chang:2024ysp}%
	\BibitemOpen
	\bibfield  {author} {\bibinfo {author} {\bibfnamefont {W.-X.}\ \bibnamefont
			{Chang}}, \bibinfo {author} {\bibfnamefont {S.}~\bibnamefont {Yin}}, \bibinfo
		{author} {\bibfnamefont {S.-X.}\ \bibnamefont {Zhang}},\ and\ \bibinfo
		{author} {\bibfnamefont {Z.-X.}\ \bibnamefont {Li}},\ }\bibfield  {title}
	{\bibinfo {title} {{Imaginary-Time Mpemba Effect in Quantum Many-Body
				Systems}},\ }\href {https://doi.org/10.1103/dvrt-n2hq} {\bibfield  {journal}
		{\bibinfo  {journal} {Phys. Rev. Lett.}\ }\textbf {\bibinfo {volume} {136}},\
		\bibinfo {pages} {100403} (\bibinfo {year} {2026})},\ \Eprint
	{https://arxiv.org/abs/2409.06547} {arXiv:2409.06547 [cond-mat.str-el]}
	\BibitemShut {NoStop}%
	\bibitem [{\citenamefont {Yu}\ \emph {et~al.}(2026)\citenamefont {Yu},
		\citenamefont {Hu},\ and\ \citenamefont {Zhang}}]{Yu:2025bgy}%
	\BibitemOpen
	\bibfield  {author} {\bibinfo {author} {\bibfnamefont {H.}~\bibnamefont
			{Yu}}, \bibinfo {author} {\bibfnamefont {J.}~\bibnamefont {Hu}},\ and\
		\bibinfo {author} {\bibfnamefont {S.-X.}\ \bibnamefont {Zhang}},\ }\bibfield
	{title} {\bibinfo {title} {{Quantum pontus-Mpemba effects in real- and
				imaginary-time dynamics}},\ }\href {https://doi.org/10.1103/4tgf-9hmc}
	{\bibfield  {journal} {\bibinfo  {journal} {Phys. Rev. B}\ }\textbf {\bibinfo
			{volume} {113}},\ \bibinfo {pages} {134304} (\bibinfo {year} {2026})},\
	\Eprint {https://arxiv.org/abs/2509.01960} {arXiv:2509.01960 [quant-ph]}
	\BibitemShut {NoStop}%
	\bibitem [{\citenamefont {Ares}\ \emph {et~al.}(2023)\citenamefont {Ares},
		\citenamefont {Murciano},\ and\ \citenamefont {Calabrese}}]{Ares:2022koq}%
	\BibitemOpen
	\bibfield  {author} {\bibinfo {author} {\bibfnamefont {F.}~\bibnamefont
			{Ares}}, \bibinfo {author} {\bibfnamefont {S.}~\bibnamefont {Murciano}},\
		and\ \bibinfo {author} {\bibfnamefont {P.}~\bibnamefont {Calabrese}},\
	}\bibfield  {title} {\bibinfo {title} {{Entanglement asymmetry as a probe of
				symmetry breaking}},\ }\href {https://doi.org/10.1038/s41467-023-37747-8}
	{\bibfield  {journal} {\bibinfo  {journal} {Nature Commun.}\ }\textbf
		{\bibinfo {volume} {14}},\ \bibinfo {pages} {2036} (\bibinfo {year}
		{2023})},\ \Eprint {https://arxiv.org/abs/2207.14693} {arXiv:2207.14693
		[cond-mat.stat-mech]} \BibitemShut {NoStop}%
	\bibitem [{\citenamefont {Yu}\ \emph {et~al.}(2025{\natexlab{b}})\citenamefont
		{Yu}, \citenamefont {Li},\ and\ \citenamefont {Zhang}}]{Yu:2025lku}%
	\BibitemOpen
	\bibfield  {author} {\bibinfo {author} {\bibfnamefont {H.}~\bibnamefont
			{Yu}}, \bibinfo {author} {\bibfnamefont {Z.-X.}\ \bibnamefont {Li}},\ and\
		\bibinfo {author} {\bibfnamefont {S.-X.}\ \bibnamefont {Zhang}},\ }\bibfield
	{title} {\bibinfo {title} {{Symmetry Breaking Dynamics in Quantum Many-Body
				Systems}},\ }\href {https://doi.org/10.1088/0256-307X/42/11/110602}
	{\bibfield  {journal} {\bibinfo  {journal} {Chin. Phys. Lett.}\ }\textbf
		{\bibinfo {volume} {42}},\ \bibinfo {pages} {110602} (\bibinfo {year}
		{2025}{\natexlab{b}})},\ \Eprint {https://arxiv.org/abs/2501.13459}
	{arXiv:2501.13459 [quant-ph]} \BibitemShut {NoStop}%
	\bibitem [{\citenamefont {Klobas}\ \emph {et~al.}(2025)\citenamefont {Klobas},
		\citenamefont {Rylands},\ and\ \citenamefont {Bertini}}]{Klobas:2024png}%
	\BibitemOpen
	\bibfield  {author} {\bibinfo {author} {\bibfnamefont {K.}~\bibnamefont
			{Klobas}}, \bibinfo {author} {\bibfnamefont {C.}~\bibnamefont {Rylands}},\
		and\ \bibinfo {author} {\bibfnamefont {B.}~\bibnamefont {Bertini}},\
	}\bibfield  {title} {\bibinfo {title} {{Translation symmetry restoration
				under random unitary dynamics}},\ }\href
	{https://doi.org/10.1103/PhysRevB.111.L140304} {\bibfield  {journal}
		{\bibinfo  {journal} {Phys. Rev. B}\ }\textbf {\bibinfo {volume} {111}},\
		\bibinfo {pages} {L140304} (\bibinfo {year} {2025})},\ \Eprint
	{https://arxiv.org/abs/2406.04296} {arXiv:2406.04296 [cond-mat.stat-mech]}
	\BibitemShut {NoStop}%
	\bibitem [{\citenamefont {Summer}\ \emph {et~al.}(2026)\citenamefont {Summer},
		\citenamefont {Moroder}, \citenamefont {Bettmann}, \citenamefont {Turkeshi},
		\citenamefont {Marvian},\ and\ \citenamefont {Goold}}]{Summer:2025wsa}%
	\BibitemOpen
	\bibfield  {author} {\bibinfo {author} {\bibfnamefont {A.}~\bibnamefont
			{Summer}}, \bibinfo {author} {\bibfnamefont {M.}~\bibnamefont {Moroder}},
		\bibinfo {author} {\bibfnamefont {L.~P.}\ \bibnamefont {Bettmann}}, \bibinfo
		{author} {\bibfnamefont {X.}~\bibnamefont {Turkeshi}}, \bibinfo {author}
		{\bibfnamefont {I.}~\bibnamefont {Marvian}},\ and\ \bibinfo {author}
		{\bibfnamefont {J.}~\bibnamefont {Goold}},\ }\bibfield  {title} {\bibinfo
		{title} {{Resource-Theoretical Unification of Mpemba Effects: Classical and
				Quantum}},\ }\href {https://doi.org/10.1103/rbt4-psfd} {\bibfield  {journal}
		{\bibinfo  {journal} {Phys. Rev. X}\ }\textbf {\bibinfo {volume} {16}},\
		\bibinfo {pages} {011065} (\bibinfo {year} {2026})},\ \Eprint
	{https://arxiv.org/abs/2507.16976} {arXiv:2507.16976 [quant-ph]} \BibitemShut
	{NoStop}%
	\bibitem [{\citenamefont {Liu}\ \emph {et~al.}(2025)\citenamefont {Liu},
		\citenamefont {Zhang}, \citenamefont {Yin}, \citenamefont {Zhang},\ and\
		\citenamefont {Yao}}]{Liu:2024uqf}%
	\BibitemOpen
	\bibfield  {author} {\bibinfo {author} {\bibfnamefont {S.}~\bibnamefont
			{Liu}}, \bibinfo {author} {\bibfnamefont {H.-K.}\ \bibnamefont {Zhang}},
		\bibinfo {author} {\bibfnamefont {S.}~\bibnamefont {Yin}}, \bibinfo {author}
		{\bibfnamefont {S.-X.}\ \bibnamefont {Zhang}},\ and\ \bibinfo {author}
		{\bibfnamefont {H.}~\bibnamefont {Yao}},\ }\bibfield  {title} {\bibinfo
		{title} {{Symmetry restoration and quantum Mpemba effect in many-body
				localization systems}},\ }\href {https://doi.org/10.1016/j.scib.2025.10.017}
	{\bibfield  {journal} {\bibinfo  {journal} {Sci. Bull.}\ }\textbf {\bibinfo
			{volume} {70}},\ \bibinfo {pages} {3991} (\bibinfo {year} {2025})},\ \Eprint
	{https://arxiv.org/abs/2408.07750} {arXiv:2408.07750 [cond-mat.dis-nn]}
	\BibitemShut {NoStop}%
	\bibitem [{\citenamefont {Foligno}\ \emph {et~al.}(2025)\citenamefont
		{Foligno}, \citenamefont {Calabrese},\ and\ \citenamefont
		{Bertini}}]{Foligno:2024jpq}%
	\BibitemOpen
	\bibfield  {author} {\bibinfo {author} {\bibfnamefont {A.}~\bibnamefont
			{Foligno}}, \bibinfo {author} {\bibfnamefont {P.}~\bibnamefont {Calabrese}},\
		and\ \bibinfo {author} {\bibfnamefont {B.}~\bibnamefont {Bertini}},\
	}\bibfield  {title} {\bibinfo {title} {{Nonequilibrium Dynamics of Charged
				Dual-Unitary Circuits}},\ }\href
	{https://doi.org/10.1103/PRXQuantum.6.010324} {\bibfield  {journal} {\bibinfo
			{journal} {PRX Quantum}\ }\textbf {\bibinfo {volume} {6}},\ \bibinfo {pages}
		{010324} (\bibinfo {year} {2025})},\ \Eprint
	{https://arxiv.org/abs/2407.21786} {arXiv:2407.21786 [cond-mat.stat-mech]}
	\BibitemShut {NoStop}%
	\bibitem [{\citenamefont {Bertini}\ \emph {et~al.}(2023)\citenamefont
		{Bertini}, \citenamefont {Calabrese}, \citenamefont {Collura}, \citenamefont
		{Klobas},\ and\ \citenamefont {Rylands}}]{Bertini:2022srv}%
	\BibitemOpen
	\bibfield  {author} {\bibinfo {author} {\bibfnamefont {B.}~\bibnamefont
			{Bertini}}, \bibinfo {author} {\bibfnamefont {P.}~\bibnamefont {Calabrese}},
		\bibinfo {author} {\bibfnamefont {M.}~\bibnamefont {Collura}}, \bibinfo
		{author} {\bibfnamefont {K.}~\bibnamefont {Klobas}},\ and\ \bibinfo {author}
		{\bibfnamefont {C.}~\bibnamefont {Rylands}},\ }\bibfield  {title} {\bibinfo
		{title} {{Nonequilibrium Full Counting Statistics and Symmetry-Resolved
				Entanglement from Space-Time Duality}},\ }\href
	{https://doi.org/10.1103/PhysRevLett.131.140401} {\bibfield  {journal}
		{\bibinfo  {journal} {Phys. Rev. Lett.}\ }\textbf {\bibinfo {volume} {131}},\
		\bibinfo {pages} {140401} (\bibinfo {year} {2023})},\ \Eprint
	{https://arxiv.org/abs/2212.06188} {arXiv:2212.06188 [cond-mat.stat-mech]}
	\BibitemShut {NoStop}%
	\bibitem [{\citenamefont {Yamashika}\ \emph {et~al.}(2024)\citenamefont
		{Yamashika}, \citenamefont {Ares},\ and\ \citenamefont
		{Calabrese}}]{Yamashika:2024hpr}%
	\BibitemOpen
	\bibfield  {author} {\bibinfo {author} {\bibfnamefont {S.}~\bibnamefont
			{Yamashika}}, \bibinfo {author} {\bibfnamefont {F.}~\bibnamefont {Ares}},\
		and\ \bibinfo {author} {\bibfnamefont {P.}~\bibnamefont {Calabrese}},\
	}\bibfield  {title} {\bibinfo {title} {{Entanglement asymmetry and quantum
				Mpemba effect in two-dimensional free-fermion systems}},\ }\href
	{https://doi.org/10.1103/PhysRevB.110.085126} {\bibfield  {journal} {\bibinfo
			{journal} {Phys. Rev. B}\ }\textbf {\bibinfo {volume} {110}},\ \bibinfo
		{pages} {085126} (\bibinfo {year} {2024})},\ \Eprint
	{https://arxiv.org/abs/2403.04486} {arXiv:2403.04486 [cond-mat.stat-mech]}
	\BibitemShut {NoStop}%
	\bibitem [{\citenamefont {Murciano}\ \emph {et~al.}(2024)\citenamefont
		{Murciano}, \citenamefont {Ares}, \citenamefont {Klich},\ and\ \citenamefont
		{Calabrese}}]{Murciano:2023qrv}%
	\BibitemOpen
	\bibfield  {author} {\bibinfo {author} {\bibfnamefont {S.}~\bibnamefont
			{Murciano}}, \bibinfo {author} {\bibfnamefont {F.}~\bibnamefont {Ares}},
		\bibinfo {author} {\bibfnamefont {I.}~\bibnamefont {Klich}},\ and\ \bibinfo
		{author} {\bibfnamefont {P.}~\bibnamefont {Calabrese}},\ }\bibfield  {title}
	{\bibinfo {title} {{Entanglement asymmetry and quantum Mpemba effect in the
				XY spin chain}},\ }\href {https://doi.org/10.1088/1742-5468/ad17b4}
	{\bibfield  {journal} {\bibinfo  {journal} {J. Stat. Mech.}\ }\textbf
		{\bibinfo {volume} {2401}},\ \bibinfo {pages} {013103} (\bibinfo {year}
		{2024})},\ \Eprint {https://arxiv.org/abs/2310.07513} {arXiv:2310.07513
		[cond-mat.stat-mech]} \BibitemShut {NoStop}%
	\bibitem [{\citenamefont {Yu}\ \emph {et~al.}(2025{\natexlab{c}})\citenamefont
		{Yu}, \citenamefont {Jin}, \citenamefont {Zhang}, \citenamefont {Xu},\ and\
		\citenamefont {Fan}}]{Yu:2025omg}%
	\BibitemOpen
	\bibfield  {author} {\bibinfo {author} {\bibfnamefont {Y.}~\bibnamefont
			{Yu}}, \bibinfo {author} {\bibfnamefont {T.}~\bibnamefont {Jin}}, \bibinfo
		{author} {\bibfnamefont {L.}~\bibnamefont {Zhang}}, \bibinfo {author}
		{\bibfnamefont {K.}~\bibnamefont {Xu}},\ and\ \bibinfo {author}
		{\bibfnamefont {H.}~\bibnamefont {Fan}},\ }\bibfield  {title} {\bibinfo
		{title} {{Tuning the quantum Mpemba effect in an isolated system by
				initial-state engineering}},\ }\href {https://doi.org/10.1103/yzjd-pk8h}
	{\bibfield  {journal} {\bibinfo  {journal} {Phys. Rev. B}\ }\textbf {\bibinfo
			{volume} {112}},\ \bibinfo {pages} {094315} (\bibinfo {year}
		{2025}{\natexlab{c}})},\ \Eprint {https://arxiv.org/abs/2505.02040}
	{arXiv:2505.02040 [quant-ph]} \BibitemShut {NoStop}%
	\bibitem [{\citenamefont {Rylands}\ \emph {et~al.}(2024)\citenamefont
		{Rylands}, \citenamefont {Klobas}, \citenamefont {Ares}, \citenamefont
		{Calabrese}, \citenamefont {Murciano},\ and\ \citenamefont
		{Bertini}}]{Rylands:2023yzx}%
	\BibitemOpen
	\bibfield  {author} {\bibinfo {author} {\bibfnamefont {C.}~\bibnamefont
			{Rylands}}, \bibinfo {author} {\bibfnamefont {K.}~\bibnamefont {Klobas}},
		\bibinfo {author} {\bibfnamefont {F.}~\bibnamefont {Ares}}, \bibinfo {author}
		{\bibfnamefont {P.}~\bibnamefont {Calabrese}}, \bibinfo {author}
		{\bibfnamefont {S.}~\bibnamefont {Murciano}},\ and\ \bibinfo {author}
		{\bibfnamefont {B.}~\bibnamefont {Bertini}},\ }\bibfield  {title} {\bibinfo
		{title} {{Microscopic Origin of the Quantum Mpemba Effect in Integrable
				Systems}},\ }\href {https://doi.org/10.1103/PhysRevLett.133.010401}
	{\bibfield  {journal} {\bibinfo  {journal} {Phys. Rev. Lett.}\ }\textbf
		{\bibinfo {volume} {133}},\ \bibinfo {pages} {010401} (\bibinfo {year}
		{2024})},\ \Eprint {https://arxiv.org/abs/2310.04419} {arXiv:2310.04419
		[cond-mat.stat-mech]} \BibitemShut {NoStop}%
	\bibitem [{\citenamefont {Xu}\ \emph {et~al.}(2025)\citenamefont {Xu} \emph
		{et~al.}}]{Xu:2025wml}%
	\BibitemOpen
	\bibfield  {author} {\bibinfo {author} {\bibfnamefont {Y.}~\bibnamefont {Xu}}
		\emph {et~al.},\ }\bibfield  {title} {\bibinfo {title} {{Observation and
				Modulation of the Quantum Mpemba Effect on a Superconducting Quantum
				Processor}},\ }\href@noop {} {\  (\bibinfo {year} {2025})},\ \Eprint
	{https://arxiv.org/abs/2508.07707} {arXiv:2508.07707 [quant-ph]} \BibitemShut
	{NoStop}%
	\bibitem [{\citenamefont {Rath}\ \emph {et~al.}(2023)\citenamefont {Rath},
		\citenamefont {Vitale}, \citenamefont {Murciano}, \citenamefont {Votto},
		\citenamefont {Dubail}, \citenamefont {Kueng}, \citenamefont {Branciard},
		\citenamefont {Calabrese},\ and\ \citenamefont {Vermersch}}]{Rath:2022qif}%
	\BibitemOpen
	\bibfield  {author} {\bibinfo {author} {\bibfnamefont {A.}~\bibnamefont
			{Rath}}, \bibinfo {author} {\bibfnamefont {V.}~\bibnamefont {Vitale}},
		\bibinfo {author} {\bibfnamefont {S.}~\bibnamefont {Murciano}}, \bibinfo
		{author} {\bibfnamefont {M.}~\bibnamefont {Votto}}, \bibinfo {author}
		{\bibfnamefont {J.}~\bibnamefont {Dubail}}, \bibinfo {author} {\bibfnamefont
			{R.}~\bibnamefont {Kueng}}, \bibinfo {author} {\bibfnamefont
			{C.}~\bibnamefont {Branciard}}, \bibinfo {author} {\bibfnamefont
			{P.}~\bibnamefont {Calabrese}},\ and\ \bibinfo {author} {\bibfnamefont
			{B.}~\bibnamefont {Vermersch}},\ }\bibfield  {title} {\bibinfo {title}
		{{Entanglement Barrier and its Symmetry Resolution: Theory and Experimental
				Observation}},\ }\href {https://doi.org/10.1103/PRXQuantum.4.010318}
	{\bibfield  {journal} {\bibinfo  {journal} {PRX Quantum}\ }\textbf {\bibinfo
			{volume} {4}},\ \bibinfo {pages} {010318} (\bibinfo {year} {2023})},\ \Eprint
	{https://arxiv.org/abs/2209.04393} {arXiv:2209.04393 [quant-ph]} \BibitemShut
	{NoStop}%
	\bibitem [{\citenamefont {Joshi}\ \emph {et~al.}(2024)\citenamefont {Joshi}
		\emph {et~al.}}]{Joshi:2024sup}%
	\BibitemOpen
	\bibfield  {author} {\bibinfo {author} {\bibfnamefont {L.~K.}\ \bibnamefont
			{Joshi}} \emph {et~al.},\ }\bibfield  {title} {\bibinfo {title} {{Observing
				the Quantum Mpemba Effect in Quantum Simulations}},\ }\href
	{https://doi.org/10.1103/PhysRevLett.133.010402} {\bibfield  {journal}
		{\bibinfo  {journal} {Phys. Rev. Lett.}\ }\textbf {\bibinfo {volume} {133}},\
		\bibinfo {pages} {010402} (\bibinfo {year} {2024})},\ \Eprint
	{https://arxiv.org/abs/2401.04270} {arXiv:2401.04270 [quant-ph]} \BibitemShut
	{NoStop}%
	\bibitem [{\citenamefont {Turkeshi}\ \emph {et~al.}(2025)\citenamefont
		{Turkeshi}, \citenamefont {Calabrese},\ and\ \citenamefont
		{De~Luca}}]{Turkeshi:2024juo}%
	\BibitemOpen
	\bibfield  {author} {\bibinfo {author} {\bibfnamefont {X.}~\bibnamefont
			{Turkeshi}}, \bibinfo {author} {\bibfnamefont {P.}~\bibnamefont
			{Calabrese}},\ and\ \bibinfo {author} {\bibfnamefont {A.}~\bibnamefont
			{De~Luca}},\ }\bibfield  {title} {\bibinfo {title} {{Quantum Mpemba Effect in
				Random Circuits}},\ }\href {https://doi.org/10.1103/5d6p-8d1b} {\bibfield
		{journal} {\bibinfo  {journal} {Phys. Rev. Lett.}\ }\textbf {\bibinfo
			{volume} {135}},\ \bibinfo {pages} {040403} (\bibinfo {year} {2025})},\
	\Eprint {https://arxiv.org/abs/2405.14514} {arXiv:2405.14514 [quant-ph]}
	\BibitemShut {NoStop}%
	\bibitem [{\citenamefont {Liu}\ \emph {et~al.}(2024)\citenamefont {Liu},
		\citenamefont {Zhang}, \citenamefont {Yin},\ and\ \citenamefont
		{Zhang}}]{Liu:2024kzv}%
	\BibitemOpen
	\bibfield  {author} {\bibinfo {author} {\bibfnamefont {S.}~\bibnamefont
			{Liu}}, \bibinfo {author} {\bibfnamefont {H.-K.}\ \bibnamefont {Zhang}},
		\bibinfo {author} {\bibfnamefont {S.}~\bibnamefont {Yin}},\ and\ \bibinfo
		{author} {\bibfnamefont {S.-X.}\ \bibnamefont {Zhang}},\ }\bibfield  {title}
	{\bibinfo {title} {{Symmetry Restoration and Quantum Mpemba Effect in
				Symmetric Random Circuits}},\ }\href
	{https://doi.org/10.1103/PhysRevLett.133.140405} {\bibfield  {journal}
		{\bibinfo  {journal} {Phys. Rev. Lett.}\ }\textbf {\bibinfo {volume} {133}},\
		\bibinfo {pages} {140405} (\bibinfo {year} {2024})},\ \Eprint
	{https://arxiv.org/abs/2403.08459} {arXiv:2403.08459 [quant-ph]} \BibitemShut
	{NoStop}%
	\bibitem [{\citenamefont {Yamashika}\ and\ \citenamefont
		{Ares}(2026)}]{Yamashika:2025vpd}%
	\BibitemOpen
	\bibfield  {author} {\bibinfo {author} {\bibfnamefont {S.}~\bibnamefont
			{Yamashika}}\ and\ \bibinfo {author} {\bibfnamefont {F.}~\bibnamefont
			{Ares}},\ }\bibfield  {title} {\bibinfo {title} {{Quantum Mpemba Effect in
				Long-Range Spin Systems}},\ }\href {https://doi.org/10.1103/52y5-8kl2}
	{\bibfield  {journal} {\bibinfo  {journal} {Phys. Rev. Lett.}\ }\textbf
		{\bibinfo {volume} {136}},\ \bibinfo {pages} {090402} (\bibinfo {year}
		{2026})},\ \Eprint {https://arxiv.org/abs/2507.06636} {arXiv:2507.06636
		[cond-mat.stat-mech]} \BibitemShut {NoStop}%
	\bibitem [{\citenamefont {Medina}\ \emph {et~al.}(2025)\citenamefont {Medina},
		\citenamefont {Culhane}, \citenamefont {Binder}, \citenamefont {Landi},\ and\
		\citenamefont {Goold}}]{Medina:2024yep}%
	\BibitemOpen
	\bibfield  {author} {\bibinfo {author} {\bibfnamefont {I.}~\bibnamefont
			{Medina}}, \bibinfo {author} {\bibfnamefont {O.}~\bibnamefont {Culhane}},
		\bibinfo {author} {\bibfnamefont {F.~C.}\ \bibnamefont {Binder}}, \bibinfo
		{author} {\bibfnamefont {G.~T.}\ \bibnamefont {Landi}},\ and\ \bibinfo
		{author} {\bibfnamefont {J.}~\bibnamefont {Goold}},\ }\bibfield  {title}
	{\bibinfo {title} {{Anomalous Discharging of Quantum Batteries: The
				Ergotropic Mpemba Effect}},\ }\href
	{https://doi.org/10.1103/PhysRevLett.134.220402} {\bibfield  {journal}
		{\bibinfo  {journal} {Phys. Rev. Lett.}\ }\textbf {\bibinfo {volume} {134}},\
		\bibinfo {pages} {220402} (\bibinfo {year} {2025})},\ \Eprint
	{https://arxiv.org/abs/2412.13259} {arXiv:2412.13259 [quant-ph]} \BibitemShut
	{NoStop}%
	\bibitem [{\citenamefont {Sapui}\ \emph {et~al.}(2026)\citenamefont {Sapui},
		\citenamefont {Konar},\ and\ \citenamefont {De}}]{Sapui:2026nle}%
	\BibitemOpen
	\bibfield  {author} {\bibinfo {author} {\bibfnamefont {T.}~\bibnamefont
			{Sapui}}, \bibinfo {author} {\bibfnamefont {T.~K.}\ \bibnamefont {Konar}},\
		and\ \bibinfo {author} {\bibfnamefont {A.~S.}\ \bibnamefont {De}},\
	}\bibfield  {title} {\bibinfo {title} {{Ergotropic Mpemba crossings in
				finite-dimensional quantum batteries}},\ }\href@noop {} {\  (\bibinfo {year}
		{2026})},\ \Eprint {https://arxiv.org/abs/2602.11056} {arXiv:2602.11056
		[quant-ph]} \BibitemShut {NoStop}%
	\bibitem [{\citenamefont {Li}\ \emph {et~al.}(2025{\natexlab{a}})\citenamefont
		{Li}, \citenamefont {Li},\ and\ \citenamefont {Li}}]{5xrr-x2rm}%
	\BibitemOpen
	\bibfield  {author} {\bibinfo {author} {\bibfnamefont {Y.}~\bibnamefont
			{Li}}, \bibinfo {author} {\bibfnamefont {W.}~\bibnamefont {Li}},\ and\
		\bibinfo {author} {\bibfnamefont {X.}~\bibnamefont {Li}},\ }\bibfield
	{title} {\bibinfo {title} {Ergotropic mpemba effect in non-markovian quantum
			systems},\ }\href {https://doi.org/10.1103/5xrr-x2rm} {\bibfield  {journal}
		{\bibinfo  {journal} {Phys. Rev. A}\ }\textbf {\bibinfo {volume} {112}},\
		\bibinfo {pages} {032209} (\bibinfo {year} {2025}{\natexlab{a}})}\BibitemShut
	{NoStop}%
	\bibitem [{\citenamefont {Yu}\ and\ \citenamefont {Eberly}(2006)}]{Yu:2006iux}%
	\BibitemOpen
	\bibfield  {author} {\bibinfo {author} {\bibfnamefont {T.}~\bibnamefont
			{Yu}}\ and\ \bibinfo {author} {\bibfnamefont {J.~H.}\ \bibnamefont
			{Eberly}},\ }\bibfield  {title} {\bibinfo {title} {{Sudden death of
				entanglement: Classical noise effects}},\ }\href
	{https://doi.org/10.1016/j.optcom.2006.01.061} {\bibfield  {journal}
		{\bibinfo  {journal} {Opt. Commun.}\ }\textbf {\bibinfo {volume} {264}},\
		\bibinfo {pages} {393} (\bibinfo {year} {2006})},\ \Eprint
	{https://arxiv.org/abs/quant-ph/0602196} {arXiv:quant-ph/0602196}
	\BibitemShut {NoStop}%
	\bibitem [{\citenamefont {Yu}\ and\ \citenamefont {Eberly}(2004)}]{Yu:2004cga}%
	\BibitemOpen
	\bibfield  {author} {\bibinfo {author} {\bibfnamefont {T.}~\bibnamefont
			{Yu}}\ and\ \bibinfo {author} {\bibfnamefont {J.~H.}\ \bibnamefont
			{Eberly}},\ }\bibfield  {title} {\bibinfo {title} {{Finite-Time
				Disentanglement Via Spontaneous Emission}},\ }\href
	{https://doi.org/10.1103/PhysRevLett.93.140404} {\bibfield  {journal}
		{\bibinfo  {journal} {Phys. Rev. Lett.}\ }\textbf {\bibinfo {volume} {93}},\
		\bibinfo {pages} {140404} (\bibinfo {year} {2004})},\ \Eprint
	{https://arxiv.org/abs/quant-ph/0404161} {arXiv:quant-ph/0404161}
	\BibitemShut {NoStop}%
	\bibitem [{\citenamefont {Almeida}\ \emph {et~al.}(2007)\citenamefont
		{Almeida}, \citenamefont {de~Melo}, \citenamefont {Hor-Meyll}, \citenamefont
		{Salles}, \citenamefont {Walborn}, \citenamefont {Ribeiro},\ and\
		\citenamefont {Davidovich}}]{Almeida:2007jib}%
	\BibitemOpen
	\bibfield  {author} {\bibinfo {author} {\bibfnamefont {M.~P.}\ \bibnamefont
			{Almeida}}, \bibinfo {author} {\bibfnamefont {F.}~\bibnamefont {de~Melo}},
		\bibinfo {author} {\bibfnamefont {M.}~\bibnamefont {Hor-Meyll}}, \bibinfo
		{author} {\bibfnamefont {A.}~\bibnamefont {Salles}}, \bibinfo {author}
		{\bibfnamefont {S.~P.}\ \bibnamefont {Walborn}}, \bibinfo {author}
		{\bibfnamefont {P.~H.~S.}\ \bibnamefont {Ribeiro}},\ and\ \bibinfo {author}
		{\bibfnamefont {L.}~\bibnamefont {Davidovich}},\ }\bibfield  {title}
	{\bibinfo {title} {{Environment-Induced Sudden Death of Entanglement}},\
	}\href {https://doi.org/10.1126/science.1139892} {\bibfield  {journal}
		{\bibinfo  {journal} {Science}\ }\textbf {\bibinfo {volume} {316}},\ \bibinfo
		{pages} {1139892} (\bibinfo {year} {2007})},\ \Eprint
	{https://arxiv.org/abs/quant-ph/0701184} {arXiv:quant-ph/0701184}
	\BibitemShut {NoStop}%
	\bibitem [{\citenamefont {Li}\ \emph {et~al.}(2025{\natexlab{b}})\citenamefont
		{Li}, \citenamefont {Shang},\ and\ \citenamefont {Wu}}]{Li:2025bzd}%
	\BibitemOpen
	\bibfield  {author} {\bibinfo {author} {\bibfnamefont {S.-H.}\ \bibnamefont
			{Li}}, \bibinfo {author} {\bibfnamefont {S.-H.}\ \bibnamefont {Shang}},\ and\
		\bibinfo {author} {\bibfnamefont {S.-M.}\ \bibnamefont {Wu}},\ }\bibfield
	{title} {\bibinfo {title} {{Does acceleration always degrade quantum
				entanglement for tetrapartite Unruh-DeWitt detectors?}},\ }\href
	{https://doi.org/10.1007/JHEP05(2025)214} {\bibfield  {journal} {\bibinfo
			{journal} {JHEP}\ }\textbf {\bibinfo {volume} {05}},\ \bibinfo {pages}
		{214}},\ \Eprint {https://arxiv.org/abs/2502.05881} {arXiv:2502.05881
		[gr-qc]} \BibitemShut {NoStop}%
	\bibitem [{\citenamefont {Liu}\ \emph {et~al.}(2026)\citenamefont {Liu},
		\citenamefont {Liu},\ and\ \citenamefont {Wu}}]{Liu:2025hcx}%
	\BibitemOpen
	\bibfield  {author} {\bibinfo {author} {\bibfnamefont {X.}~\bibnamefont
			{Liu}}, \bibinfo {author} {\bibfnamefont {W.}~\bibnamefont {Liu}},\ and\
		\bibinfo {author} {\bibfnamefont {S.-M.}\ \bibnamefont {Wu}},\ }\bibfield
	{title} {\bibinfo {title} {{Entanglement degradation of static black holes in
				effective quantum gravity}},\ }\href
	{https://doi.org/10.1016/j.physletb.2026.140334} {\bibfield  {journal}
		{\bibinfo  {journal} {Phys. Lett. B}\ }\textbf {\bibinfo {volume} {875}},\
		\bibinfo {pages} {140334} (\bibinfo {year} {2026})},\ \Eprint
	{https://arxiv.org/abs/2511.12245} {arXiv:2511.12245 [gr-qc]} \BibitemShut
	{NoStop}%
	\bibitem [{\citenamefont {Gallock-Yoshimura}\ and\ \citenamefont
		{Mann}(2021)}]{Gallock-Yoshimura:2021xsy}%
	\BibitemOpen
	\bibfield  {author} {\bibinfo {author} {\bibfnamefont {K.}~\bibnamefont
			{Gallock-Yoshimura}}\ and\ \bibinfo {author} {\bibfnamefont {R.~B.}\
			\bibnamefont {Mann}},\ }\bibfield  {title} {\bibinfo {title} {{Entangled
				detectors nonperturbatively harvest mutual information}},\ }\href
	{https://doi.org/10.1103/PhysRevD.104.125017} {\bibfield  {journal} {\bibinfo
			{journal} {Phys. Rev. D}\ }\textbf {\bibinfo {volume} {104}},\ \bibinfo
		{pages} {125017} (\bibinfo {year} {2021})},\ \Eprint
	{https://arxiv.org/abs/2109.07495} {arXiv:2109.07495 [quant-ph]} \BibitemShut
	{NoStop}%
	\bibitem [{\citenamefont {Bellomo}\ \emph {et~al.}(2008)\citenamefont
		{Bellomo}, \citenamefont {Franco},\ and\ \citenamefont
		{Compagno}}]{Bellomo:2007mms}%
	\BibitemOpen
	\bibfield  {author} {\bibinfo {author} {\bibfnamefont {B.}~\bibnamefont
			{Bellomo}}, \bibinfo {author} {\bibfnamefont {R.~L.}\ \bibnamefont
			{Franco}},\ and\ \bibinfo {author} {\bibfnamefont {G.}~\bibnamefont
			{Compagno}},\ }\bibfield  {title} {\bibinfo {title} {{Entanglement dynamics
				of two independent qubits in environments with and without memory}},\ }\href
	{https://doi.org/10.1103/PhysRevA.77.032342} {\bibfield  {journal} {\bibinfo
			{journal} {Phys. Rev. A}\ }\textbf {\bibinfo {volume} {77}},\ \bibinfo
		{pages} {032342} (\bibinfo {year} {2008})},\ \Eprint
	{https://arxiv.org/abs/0711.4799} {arXiv:0711.4799 [quant-ph]} \BibitemShut
	{NoStop}%
	\bibitem [{\citenamefont {Bellomo}\ \emph {et~al.}(2007)\citenamefont
		{Bellomo}, \citenamefont {Franco},\ and\ \citenamefont
		{Compagno}}]{Bellomo:2007euv}%
	\BibitemOpen
	\bibfield  {author} {\bibinfo {author} {\bibfnamefont {B.}~\bibnamefont
			{Bellomo}}, \bibinfo {author} {\bibfnamefont {R.~L.}\ \bibnamefont
			{Franco}},\ and\ \bibinfo {author} {\bibfnamefont {G.}~\bibnamefont
			{Compagno}},\ }\bibfield  {title} {\bibinfo {title} {{Non-Markovian Effects
				on the Dynamics of Entanglement}},\ }\href
	{https://doi.org/10.1103/PhysRevLett.99.160502} {\bibfield  {journal}
		{\bibinfo  {journal} {Phys. Rev. Lett.}\ }\textbf {\bibinfo {volume} {99}},\
		\bibinfo {pages} {160502} (\bibinfo {year} {2007})},\ \Eprint
	{https://arxiv.org/abs/0804.2377} {arXiv:0804.2377 [quant-ph]} \BibitemShut
	{NoStop}%
	\bibitem [{\citenamefont {Pan}\ \emph {et~al.}(2026)\citenamefont {Pan},
		\citenamefont {Zhang},\ and\ \citenamefont {Cai}}]{Pan:2025oqc}%
	\BibitemOpen
	\bibfield  {author} {\bibinfo {author} {\bibfnamefont {Y.}~\bibnamefont
			{Pan}}, \bibinfo {author} {\bibfnamefont {B.}~\bibnamefont {Zhang}},\ and\
		\bibinfo {author} {\bibfnamefont {Q.}~\bibnamefont {Cai}},\ }\bibfield
	{title} {\bibinfo {title} {{Entanglement is protected by acceleration-induced
				transparency in thermal field}},\ }\href {https://doi.org/10.1103/z6nb-gjry}
	{\bibfield  {journal} {\bibinfo  {journal} {Phys. Rev. D}\ }\textbf {\bibinfo
			{volume} {113}},\ \bibinfo {pages} {025004} (\bibinfo {year} {2026})},\
	\Eprint {https://arxiv.org/abs/2512.06043} {arXiv:2512.06043 [quant-ph]}
	\BibitemShut {NoStop}%
	\bibitem [{\citenamefont {Cavalcante}\ \emph {et~al.}(2025)\citenamefont
		{Cavalcante}, \citenamefont {Bonan{\c{c}}a}, \citenamefont {Miranda},\ and\
		\citenamefont {Deffner}}]{Cavalcante:2025wlm}%
	\BibitemOpen
	\bibfield  {author} {\bibinfo {author} {\bibfnamefont {M.~F.}\ \bibnamefont
			{Cavalcante}}, \bibinfo {author} {\bibfnamefont {M.~V.~S.}\ \bibnamefont
			{Bonan{\c{c}}a}}, \bibinfo {author} {\bibfnamefont {E.}~\bibnamefont
			{Miranda}},\ and\ \bibinfo {author} {\bibfnamefont {S.}~\bibnamefont
			{Deffner}},\ }\bibfield  {title} {\bibinfo {title} {{Emergence of X states in
				a quantum impurity model}},\ }\href
	{https://doi.org/10.1103/PhysRevResearch.7.L022027} {\bibfield  {journal}
		{\bibinfo  {journal} {Phys. Rev. Res.}\ }\textbf {\bibinfo {volume} {7}},\
		\bibinfo {pages} {L022027} (\bibinfo {year} {2025})},\ \Eprint
	{https://arxiv.org/abs/2501.13914} {arXiv:2501.13914 [cond-mat.str-el]}
	\BibitemShut {NoStop}%
	\bibitem [{\citenamefont {Nakamura}\ and\ \citenamefont
		{Ankerhold}(2026)}]{Nakamura:2025mae}%
	\BibitemOpen
	\bibfield  {author} {\bibinfo {author} {\bibfnamefont {K.}~\bibnamefont
			{Nakamura}}\ and\ \bibinfo {author} {\bibfnamefont {J.}~\bibnamefont
			{Ankerhold}},\ }\bibfield  {title} {\bibinfo {title} {{Entanglement dynamics
				and performance of two-qubit gates for superconducting qubits under
				non-Markovian effects}},\ }\href {https://doi.org/10.1103/b5jp-s6t2}
	{\bibfield  {journal} {\bibinfo  {journal} {Phys. Rev. Res.}\ }\textbf
		{\bibinfo {volume} {8}},\ \bibinfo {pages} {013337} (\bibinfo {year}
		{2026})},\ \Eprint {https://arxiv.org/abs/2510.05872} {arXiv:2510.05872
		[quant-ph]} \BibitemShut {NoStop}%
	\bibitem [{\citenamefont {Rau}(2009)}]{Rau:2009ufg}%
	\BibitemOpen
	\bibfield  {author} {\bibinfo {author} {\bibfnamefont {A.~R.~P.}\
			\bibnamefont {Rau}},\ }\bibfield  {title} {\bibinfo {title} {{Algebraic
				characterization of X-states in quantum information}},\ }\href
	{https://doi.org/10.1088/1751-8113/42/41/412002} {\bibfield  {journal}
		{\bibinfo  {journal} {J. Phys. A}\ }\textbf {\bibinfo {volume} {42}},\
		\bibinfo {pages} {412002} (\bibinfo {year} {2009})},\ \Eprint
	{https://arxiv.org/abs/0906.4716} {arXiv:0906.4716 [quant-ph]} \BibitemShut
	{NoStop}%
	\bibitem [{\citenamefont {Yu}\ and\ \citenamefont {Eberly}(2007)}]{Yu:2007bwc}%
	\BibitemOpen
	\bibfield  {author} {\bibinfo {author} {\bibfnamefont {T.}~\bibnamefont
			{Yu}}\ and\ \bibinfo {author} {\bibfnamefont {J.~H.}\ \bibnamefont
			{Eberly}},\ }\bibfield  {title} {\bibinfo {title} {{Evolution from
				entanglement to decoherence of bipartite mixed ''X'' states}},\ }\href
	{https://doi.org/10.26421/QIC7.5-6-3} {\bibfield  {journal} {\bibinfo
			{journal} {Quant. Inf. Comput.}\ }\textbf {\bibinfo {volume} {7}},\ \bibinfo
		{pages} {459} (\bibinfo {year} {2007})},\ \Eprint
	{https://arxiv.org/abs/quant-ph/0503089} {arXiv:quant-ph/0503089}
	\BibitemShut {NoStop}%
	\bibitem [{\citenamefont {Hill}\ and\ \citenamefont
		{Wootters}(1997)}]{Hill:1997pfa}%
	\BibitemOpen
	\bibfield  {author} {\bibinfo {author} {\bibfnamefont {S.}~\bibnamefont
			{Hill}}\ and\ \bibinfo {author} {\bibfnamefont {W.~K.}\ \bibnamefont
			{Wootters}},\ }\bibfield  {title} {\bibinfo {title} {{Entanglement of a pair
				of quantum bits}},\ }\href {https://doi.org/10.1103/PhysRevLett.78.5022}
	{\bibfield  {journal} {\bibinfo  {journal} {Phys. Rev. Lett.}\ }\textbf
		{\bibinfo {volume} {78}},\ \bibinfo {pages} {5022} (\bibinfo {year}
		{1997})},\ \Eprint {https://arxiv.org/abs/quant-ph/9703041}
	{arXiv:quant-ph/9703041} \BibitemShut {NoStop}%
	\bibitem [{\citenamefont {Wootters}(1998)}]{Wootters:1997id}%
	\BibitemOpen
	\bibfield  {author} {\bibinfo {author} {\bibfnamefont {W.~K.}\ \bibnamefont
			{Wootters}},\ }\bibfield  {title} {\bibinfo {title} {{Entanglement of
				formation of an arbitrary state of two qubits}},\ }\href
	{https://doi.org/10.1103/PhysRevLett.80.2245} {\bibfield  {journal} {\bibinfo
			{journal} {Phys. Rev. Lett.}\ }\textbf {\bibinfo {volume} {80}},\ \bibinfo
		{pages} {2245} (\bibinfo {year} {1998})},\ \Eprint
	{https://arxiv.org/abs/quant-ph/9709029} {arXiv:quant-ph/9709029}
	\BibitemShut {NoStop}%
	\bibitem [{\citenamefont {Klich}\ \emph {et~al.}(2019)\citenamefont {Klich},
		\citenamefont {Raz}, \citenamefont {Hirschberg},\ and\ \citenamefont
		{Vucelja}}]{PhysRevX.9.021060}%
	\BibitemOpen
	\bibfield  {author} {\bibinfo {author} {\bibfnamefont {I.}~\bibnamefont
			{Klich}}, \bibinfo {author} {\bibfnamefont {O.}~\bibnamefont {Raz}}, \bibinfo
		{author} {\bibfnamefont {O.}~\bibnamefont {Hirschberg}},\ and\ \bibinfo
		{author} {\bibfnamefont {M.}~\bibnamefont {Vucelja}},\ }\bibfield  {title}
	{\bibinfo {title} {Mpemba index and anomalous relaxation},\ }\href
	{https://doi.org/10.1103/PhysRevX.9.021060} {\bibfield  {journal} {\bibinfo
			{journal} {Phys. Rev. X}\ }\textbf {\bibinfo {volume} {9}},\ \bibinfo {pages}
		{021060} (\bibinfo {year} {2019})}\BibitemShut {NoStop}%
	\bibitem [{\citenamefont {Kumar}\ \emph {et~al.}(2022)\citenamefont {Kumar},
		\citenamefont {Chétrite},\ and\ \citenamefont
		{Bechhoefer}}]{doi:10.1073/pnas.2118484119}%
	\BibitemOpen
	\bibfield  {author} {\bibinfo {author} {\bibfnamefont {A.}~\bibnamefont
			{Kumar}}, \bibinfo {author} {\bibfnamefont {R.}~\bibnamefont {Chétrite}},\
		and\ \bibinfo {author} {\bibfnamefont {J.}~\bibnamefont {Bechhoefer}},\
	}\bibfield  {title} {\bibinfo {title} {Anomalous heating in a colloidal
			system},\ }\href {https://doi.org/10.1073/pnas.2118484119} {\bibfield
		{journal} {\bibinfo  {journal} {Proc. Natl. Acad. Sci. U.S.A.}\ }\textbf
		{\bibinfo {volume} {119}},\ \bibinfo {pages} {e2118484119} (\bibinfo {year}
		{2022})}\BibitemShut {NoStop}%
\end{thebibliography}
%

\section*{Appendix}
\setcounter{figure}{0}
\renewcommand{\figurename}{Appendix Figure}
We provide the explicit derivation steps for the asymmetric dissipation framework and present supplementary analysis regarding the geometric boundaries of the phase diagrams. When $\Gamma_A \neq \Gamma_B$, the time-dependent density matrix elements take the following form:
\begin{align}
	a(t) &= c_2 + d(0)\gamma_A\gamma_B - d(0)(\gamma_A+\gamma_B), \\
	b(t) &= d(0)\gamma_B(1-\gamma_A) + b(0)\gamma_B, \\
	c(t) &= d(0)\gamma_A(1-\gamma_B) + c(0)\gamma_A, \\
	d(t) &= d(0)\gamma_A\gamma_B, \\
	z(t) &=z(0)\sqrt{\gamma_A\gamma_B},\\
	w(t) &= w(0)\sqrt{\gamma_A\gamma_B}.
\end{align}
By imposing the initial conditions $b(0) = c(0) = 0$, the integration constant $c_2$ is uniquely determined by the trace normalization condition $\text{Tr}[\rho(t)] = 1$, yielding $c_2 = 1$. The matrix elements for the symmetric case follow directly from the general expressions by setting \(\gamma_A = \gamma_B = \gamma\). Consequently, the key quantity governing the concurrence dynamics can be simplified as:
\begin{equation}
	|w(t)|-\sqrt{b(t)c(t)}=\sqrt{\gamma_A \gamma_B}\left(|w(0)| - d(0)\sqrt{(1-\gamma_A)(1-\gamma_B)}\right),
\end{equation}
which corresponds precisely to Eq.~(\ref{eq_conc}) in the main text. Importantly, the positive semi-definiteness of the density matrix requires the underlying condition
\begin{equation}
	a(t)d(t) - |w(t)|^2 = \gamma_A\gamma_B\left[d(0)-|w(0)|^2-d(0)^2 f(\gamma_A,\gamma_B)\right]\geq 0.
\end{equation}
Let us define the auxiliary function $f(x, y) = x + y - xy$, whose partial derivatives are given by: $\partial f / \partial x = 1 - y, \partial f / \partial y = 1 - x.$
For $\gamma_A,\gamma_B\in[0,1]$, the function is nondecreasing in each argument and attains its global maximum $f_{\max}=1$ whenever $\gamma_A=1$ or $\gamma_B=1$.
\begin{figure}[t]
	\centering{\includegraphics[width=0.8\linewidth]{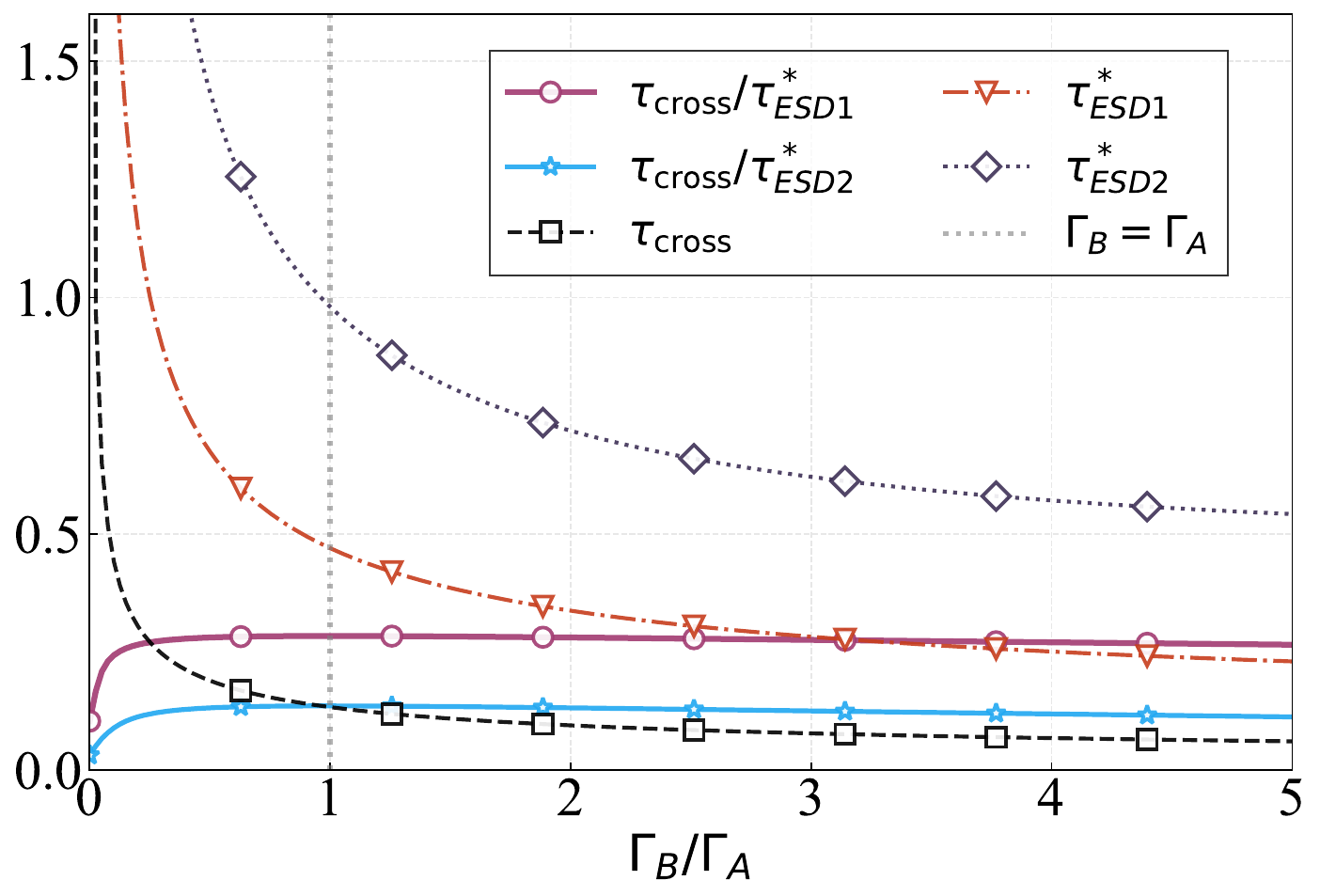}
		\caption{Characteristic timescales and their dimensionless ratios vary as the ratio $\Gamma_B/\Gamma_A$. The black dashed line (squares) represents the concurrence trajectory crossing time $\tau_{\text{cross}}$. Orange dot-dashed line (downward triangles) and purple dotted line (diamonds) show ESD times $\tau^*_{\text{ESD1}}$ and $\tau^*_{\text{ESD2}}$ for two distinct initial states. Pink solid line (circles) and blue solid line (stars) give the corresponding ratios $\tau_{\text{cross}}/\tau^*_{\text{ESD1}}$ and $\tau_{\text{cross}}/\tau^*_{\text{ESD2}}$. Vertical grey dotted line marks symmetric dissipation ($\Gamma_B = \Gamma_A$).}\label{Appendix Figure 1}
	}
\end{figure}
As a result, the physical parameter regime satisfying the positivity condition is completely decoupled from the dissipation asymmetry, remaining strictly bounded by the semicircular constraint $d(0) - |w(0)|^2 - d(0)^2 \ge 0$.Furthermore, the ESD  time $\tau^*_{\text{ESD}}$ and the concurrence trajectory crossing time $\tau_{\text{cross}}$ can be straightforwardly derived from their defining conditions, matching Eqs. (\ref{eq_esd}) and (\ref{eq_cro}) in the main text.

To elucidate why the boundaries of the Mpemba region in the phase diagram are invariant under environmental asymmetry, we illustrate the dynamical behavior of these timescales in Appendix Figure.~\ref{Appendix Figure 1}. Specifically, we plot the ESD time $\tau^*_{\text{ESD}}$, the crossing time $\tau_{\text{cross}}$ (dashed line), and their ratio $\tau_{\text{cross}}/\tau^*_{\text{ESD}}$ (solid line) as functions of the dissipation ratio $\Gamma_B/\Gamma_A$. Although both timescales decrease monotonically as the ratio increases, the ESD time consistently bounds the crossing time from above, and the two curves never intersect. This demonstrates that the trajectory crossing invariably precedes the sudden death of entanglement, a feature verified by the fact that the ratio $\tau_{\text{cross}}/\tau^*_{\text{ESD}}$ remains strictly less than unity across all parameter configurations. Consequently, the parameter boundaries defining the Entanglement Mpemba region remain remarkably robust against the introduction of dissipation asymmetry.
%


\end{document}